\documentclass[preprint2]{aastex631}
\usepackage{amsmath,amssymb,amsthm}

\theoremstyle{remark}
\newtheorem*{remark}{Remark}

\begin{document}
\title{Variability Signatures of a Burst Process in Flaring Gamma-ray Blazars}

\correspondingauthor{Aryeh Brill}
\email{aryeh.brill@nasa.gov}

\author{A.~Brill}
\affiliation{NASA Postdoctoral Program Fellow, NASA Goddard Space Flight Center, Greenbelt, MD 20771, USA}

\begin{abstract}
Blazars exhibit stochastic flux variability across the electromagnetic spectrum, often exhibiting heavy-tailed flux distributions, commonly modeled as lognormal. However, \citet{Tavecchio2020} and \citet{Adams2022} found that the high-energy gamma-ray flux distributions of several of the brightest flaring \textit{Fermi}-LAT flat spectrum radio quasars (FSRQs) are well modeled by an even heavier-tailed distribution, which we show is the \textit{inverse gamma} distribution. We propose an autoregressive inverse gamma variability model in which an inverse gamma flux distribution arises as a consequence of a shot-noise process. In this model, discrete bursts are individually unresolved and averaged over within time bins, as in the analysis of \textit{Fermi}-LAT data. Stochastic variability on timescales longer than the time bin duration is modeled using first-order autoregressive structure. The flux distribution becomes approximately lognormal in the limiting case of many weak bursts. The fractional variability is predicted to decrease as the time bin duration increases. Using simulated light curves, we show that the proposed model is consistent with the typical gamma-ray variability properties of FSRQs and BL Lac objects. The model parameters can be physically interpreted as the average burst rate, the burst fluence, and the timescale of long-term stochastic fluctuations.
\end{abstract}

\section{Introduction}

Blazars are a class of active galactic nuclei (AGN) which are thought to possess relativistic jets oriented nearly along our line of sight \citep{Urry1995}. Blazars make up the vast majority of extragalactic gamma-ray sources that have been detected in the GeV band by \textit{Fermi}-LAT \citep{Ajello2020}. 

Blazars are highly variable objects at essentially all wavelengths, with variability observed in the GeV band at timescales of years down to several minutes \citep[e.g.][]{Ackermann2016}. Observations of variability can help us understand the physical processes that give rise to the high-energy emission from these objects. Characteristic timescales of variability constrain the apparent size of the emission region by placing upper bounds on the light crossing time, and are a key input for modeling the physical processes giving rise to gamma-ray emission.

Blazars can be divided into two source classes, flat spectrum radio quasars (FSRQs) and BL Lac objects. FSRQs have greater synchrotron luminosity and lower peak synchrotron frequency than BL Lac objects. Although these two source classes are often considered to form a continuous ``blazar sequence'' \citep{Fossati1998}, they may be better understood as belonging to two intrinsically distinct AGN populations, of high-power and low-power objects, respectively \citep{Keenan2021}. In the GeV gamma-ray band, FSRQs are typically more luminous and more variable than BL Lac objects  \citep{Abdo2010}.

Blazars typically have a power spectral density (PSD) with a power-law shape, indicating that the emission can be described as a stochastic process, or random walk \citep[e.g.][]{Abdo2010}. Characteristic variability timescales may appear as features or spectral breaks in the PSD. Previous works have found such breaks (when present) at short timescales of $\sim$1 day, potentially connected to the timescale of successive flare events giving rise to the high-energy emission, and at long timescales of $\sim$1 year, which may be related to processes in the accretion disk \citep[e.g.][]{Kataoka2001, Ryan2019, Tarnopolski2020}. However, gamma-ray blazars are also notable for undergoing intermediate-duration flaring periods which can last for many days, during which the gamma-ray flux can in some cases increase by an order of magnitude or more.

One example of the flaring phenomenon comes from the bright FSRQ 3C~279, which has undergone several such flares \citep[e.g.][]{Ackermann2016, Adams2022}. In particular, \citet{Adams2022} analyzed its sub-daily variability during 10 bright flaring periods lasting between 1 and 11 days, for a total of 54 days of observations. From modeling using exponential profiles, each flare was resolved into between 1 and 4 separate components, for a total of 24 distinct components that had rise and decay times ranging from days to less than 1 hour. The fluences of the flare components were found to have a dynamic range about an order of magnitude narrower than that of their rise and decay timescales, with a median fluence of $0.85 \times 10^{-3}$~erg~cm$^{-2}$.

Blazar variability can be usefully studied in the time domain using autoregressive processes, a powerful class of time series models well suited for describing stationary stochastic processes \citep[see][and references therein]{Scargle1981}. In an autoregressive (AR) process of order $p$, the value $X_i$ at a given time step $i$ is a linear combination of the values at the previous $p$ time steps, to which a stochastic ``innovation'' term $\epsilon$ is added. In the related moving average (MA) process of order $q$, a linear combination of the values of the innovations at the previous $q$ time steps is used instead. Combining these processes results in a so-called ARMA process of order $(p, q)$. In this paper, we restrict our attention to AR(1) processes, that is, the case $p = 1$, $q = 0$, and we will use the term ``autoregressive process'' broadly to refer to the class of ARMA processes in general.

Variability can also be studied through the flux distribution, that is, the probability density function (PDF) describing the observed flux values. Blazars are generally found to have heavy-tailed flux distributions, with excesses at high fluxes compared to a normal distribution. These distributions are commonly modeled as a lognormal distribution, which can arise through the multiplicative combination of many independent events \citep{Uttley2005}. Variability consistent with lognormality has been observed in many studies of blazars using data in multiple wavebands \citep[e.g.][]{Giebels2009, Kushwaha2017, Valverde2020, Acciari2021, Bhatta2021}.

A commonly observed feature of variable emission from AGN, first discussed in depth by \citet{Uttley2001}, is an approximately linear relationship between the mean flux and standard deviation calculated within subintervals of the light curve, known as the linear RMS-flux relation. \citet{Uttley2005} showed that a linear RMS-flux relation can be associated with lognormal variability, suggesting that the long-term memory needed to produce a linear RMS-flux relation must be connected with an underlying multiplicative process, ruling out an additive origin for variability, such as shot noise \citep{Lehto1989}. Importantly, however, \citet{Scargle2020} showed that an approximately linear RMS-flux relation arises as a general statistical property of any light curve that can be modeled as an autoregressive process, with the exact form of the relation depending on the shape of the innovation. An underlying multiplicative process is therefore not required to explain the observed properties of blazar light curves.

In this work, we propose a model in which blazar variability results from an underlying process of Poisson-distributed bursts. Critically, the shapes of the individual bursts are taken to be unresolved. The variability then arises from the statistics of the burst production process, giving rise to an inverse gamma flux distribution. We derive a model that yields an inverse gamma flux distribution with AR(1) autoregressive structure. The model has only three free parameters, representing the average rate of the putative bursts, the burst fluence, and an autocorrelation timescale representing long-term fluctuations in the burst rate. These parameters can be interpreted in terms of physical processes, such as plasmoid-powered flares in a relativistic magnetic reconnection scenario.

In Section~\ref{sec:background}, we provide theoretical background on autoregressive processes and the inverse gamma distribution, and discuss related work. In Section~\ref{sec:model}, we motivate the autoregressive inverse gamma model, derive its properties, and discuss important limiting cases. In Section~\ref{sec:simulation}, we use the proposed process to simulate light curves representative of several source classes. In Section~\ref{sec:discussion}, we discuss some of the model's limitations, extensions, and implications; connections to theoretical models; comparisons to previous work; and applications to blazar classification. Finally, we summarize our conclusions in Section~\ref{sec:conclusion}. 
\section{Background and Related Work}
\label{sec:background}

An ARMA process of order $(p, q)$ can be written as

\begin{equation}\label{eq:arma}
    X_i = \sum_{j=1}^p \phi_j X_{i - j} + \sum_{k=1}^q \theta_k \epsilon_{i - k} + \epsilon_i,
\end{equation}

\noindent where $X_i$ are the values of the time series, $\epsilon_i$ are stochastic innovation terms, and $\phi_j$, $\theta_k$ are the constant AR and MA coefficients, respectively. In this paper, we consider the special case of AR(1) processes with the form

\begin{equation}\label{eq:ar1}
    X_i = \phi X_{i - 1} + \epsilon_i.
\end{equation}

An AR(1) process in which the innovation terms are assumed to be normally distributed, such that

\begin{equation}\label{eq:normal_innovation}
    \epsilon \sim \mathcal{N}\left(\mu', \sigma^2 \right),
\end{equation}

\noindent where $\mu' \equiv (1 - \phi)\mu$ is the mean and $\sigma^2$ is the variance of a Gaussian distribution, is commonly referred to in astrophysics as an Ornstein-Uhlenbeck or damped random walk process. This process and its continuous-time analogue have found widespread application in modeling AGN light curves \citep[e.g.][]{Kelly2009, Moreno2019, Burd2021}. The continuous-time analogue of the general ARMA process, called CARMA, has been fruitfully deployed as well \citep{Kelly2014, Ryan2019}.

An AR(1) process is associated with an expected marginal distribution\footnote{The marginal distribution is identical to the expected flux distribution if the modeled time series is the flux light curve. These distributions will be different, though closely related, if a different time series is being considered, such as that of the logarithm of the flux.}. If the innovation is normally distributed according to Eq.~\ref{eq:normal_innovation}, the marginal distribution is normally distributed as well, such that

\begin{equation}\label{eq:gaussian_ar1}
    X \sim \mathcal{N}\left(\mu, \frac{\sigma^2}{1 - \phi^2}\right).
\end{equation}

A lognormal flux distribution is usually obtained from an AR(1) model with a normal innovation by fitting it with the logarithm of the flux, $X = \log F$. Conveniently, this transformation also prevents the possibility of generating nonphysical negative fluxes.

One way to interpret the flux distribution is as the result of the combination of two distinct underlying processes, represented by the autoregressive term parameterized by $\phi$ and the innovation term parameterized by $\mu$ and $\sigma$, which may in general have different physical interpretations and characteristic timescales. The innovation term contributes short-timescale variability, which is enhanced by the long-timescale variability provided by the autoregressive term. These are both sources of intrinsic variability, which are independent of any extrinsic uncertainty contributed by measurement error.

Recently, \citet{Tavecchio2020} proposed a model of blazar variability based on a stochastic differential equation (SDE) that exhibits nonlinear dynamics. Their model is physically motivated by considering a magnetically arrested accretion disk. The accumulation of magnetic energy is modeled as a deterministic, equilibrium-reverting process (the autoregressive term) and its dissipation as a stochastic process (the innovation term). In this model, the standard deviation of the innovation term depends linearly on the flux, so that its discrete representation is equivalent to Eq.~\ref{eq:ar1} now with

\begin{equation}\label{eq:tavecchio_innovation}
    \epsilon \sim \mathcal{N}\left(\mu', (\sigma X_{i - 1})^2 \right).
\end{equation}

Assuming that $X > 0$, the marginal distribution of this process is given by \citep[][Eq.~A8]{Tavecchio2020}

\begin{equation}\label{eq:tavecchio_pdf}
    f(X) = \frac{(\lambda\mu)^{1 + \lambda}}{\Gamma(1 + \lambda)} \frac{e^{-\lambda\mu/X}}{X^{\lambda + 2}},
\end{equation}

\noindent where $\lambda \equiv 2\phi/\sigma^2$ and $\Gamma(x)$ is the Gamma function\footnote{In the notation of \citet{Tavecchio2020}, $\lambda \equiv 2\theta/\sigma^2$.}. \citet{Tavecchio2020} showed that this PDF provides a satisfactory representation of the gamma-ray flux distributions of six bright FSRQs observed by \textit{Fermi}-LAT using the light curves analyzed by \citet{Meyer2019}. Subsequently, \citet{Adams2022} demonstrated that this model provides a better fit than a lognormal PDF to the \textit{Fermi}-LAT flux distributions of three bright FSRQs, 3C~279, PKS~1222+216, and Ton~599.

In fact, Eq.~\ref{eq:tavecchio_pdf} has the form of a well-known probability distribution, the inverse gamma distribution, which has the PDF

\begin{equation}\label{eq:inverse_gamma}
    f(X) = \frac{\beta^\alpha}{\Gamma(\alpha)} \frac{e^{-\beta/X}}{X^{\alpha + 1}}.
\end{equation}

It can be seen that Eq.~\ref{eq:tavecchio_pdf} is equivalent to Eq.~\ref{eq:inverse_gamma} with $\alpha = 1 + \lambda$ and $\beta = \lambda\mu$.

The inverse gamma distribution and gamma distribution are intimately related. As the names of these two distributions suggest, if $X \sim \mathrm{InvGamma}(\alpha, \beta)$, then $1/X \sim \mathrm{Gamma}(\alpha, \beta)$, and vice-versa\footnote{We will also use the notation $\Gamma(\alpha, \beta)$ with two arguments to denote the Gamma distribution, and $\Gamma(x)$ with one argument for the Gamma function.}. The PDF of the gamma distribution is given by

\begin{equation}
    f(X) = \frac{X^{\alpha - 1} e^{-\beta X} \beta^\alpha}{\Gamma(\alpha)}.
\end{equation}

\begin{figure}[htb]
    \centering
    \includegraphics[width=0.45\textwidth]{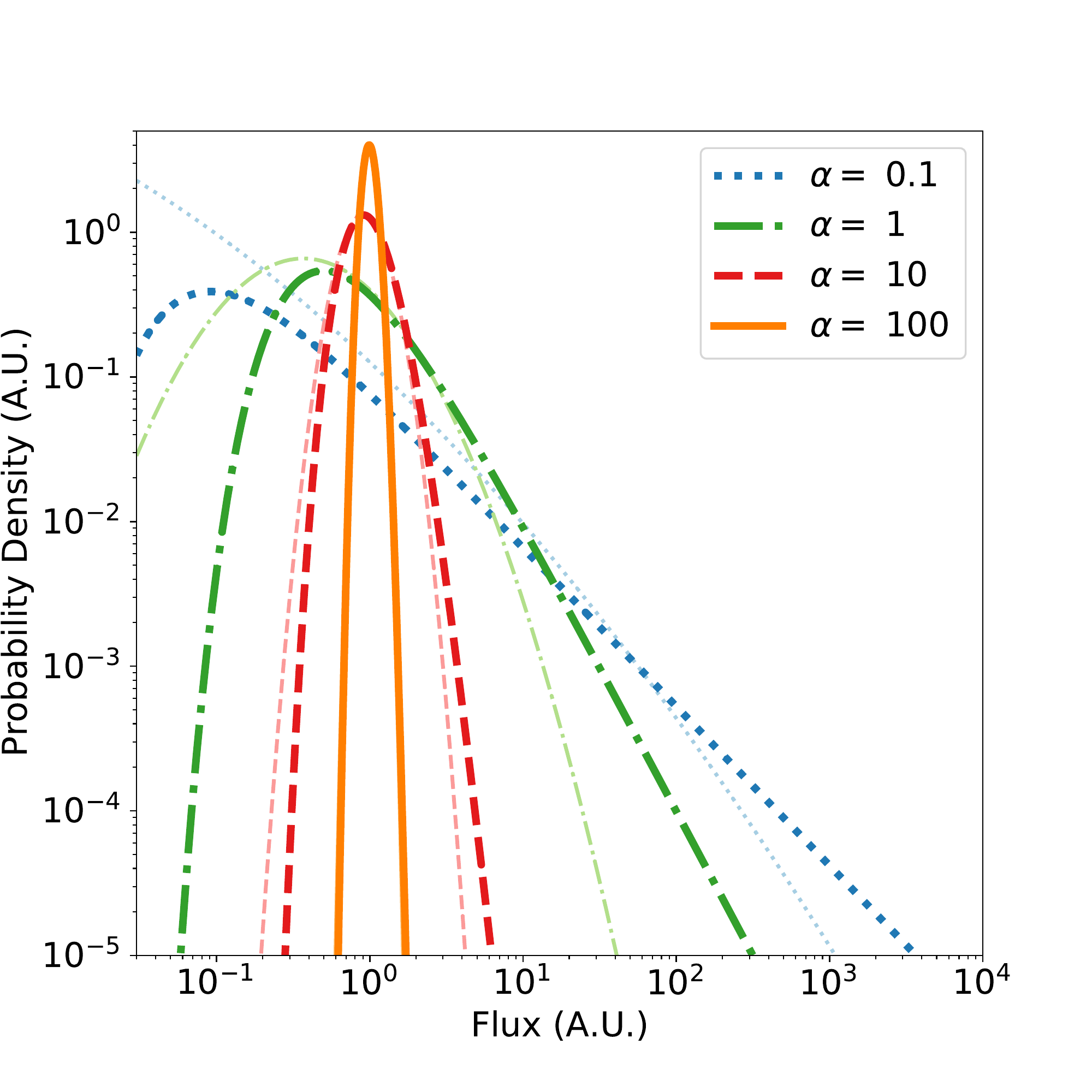}
    \caption{Inverse gamma PDFs for four values of $\alpha$, letting $\beta = \alpha$. The light thin lines show the lognormal approximation to each inverse gamma PDF determined using the results derived in Section~\ref{sec:lognormal_approximation}. For $\alpha = 100$, the inverse gamma and lognormal curves are indistinguishable on the scale of the plot.}
    \label{fig:inverse_gamma}
\end{figure}

The inverse gamma and gamma distributions have positive support, that is, $X \in (0, \infty)$. For both distributions, $\alpha > 0$ is a shape parameter determining the form of the distribution. Figure~\ref{fig:inverse_gamma} shows inverse gamma PDFs plotted for a range of $\alpha$ values. The inverse gamma distribution is highly skewed, with a high-flux tail that is heavier than that of a lognormal distribution and resembles a power law. In the regime $\alpha \gg 1$, the distribution becomes approximately (but not exactly) lognormal. The scale of the inverse gamma distribution is determined by $\beta > 0$, which for the gamma distribution is an inverse scale (or rate) parameter.

In a Poisson process, the interarrival time, or time difference between each event and the next arriving independently, is randomly distributed following an exponential distribution. The sum of $\alpha$ independent exponential variables with rate parameter $\beta$ is distributed as $\mathrm{Gamma}(\alpha, \beta)$. For this reason, the gamma distribution is closely connected with Poisson processes.

In this work, we propose an autoregressive inverse gamma model that shares several commonalities with the model of \citet{Tavecchio2020}, including an inverse gamma flux distribution and autoregressive time structure. However, the proposed model has a different physical motivation and mathematical basis deriving from an underlying emission process of discrete unresolved bursts, rather than from the nonlinear stochastic term of Eq.~\ref{eq:tavecchio_innovation}.

These differences lead to several novel predictions. In the proposed model, variability is controlled primarily by the average burst rate. The model produces isolated ``loner flares'' when the average burst rate is low; flaring light curves reminiscent of highly variable FSRQs and BL~Lac objects when it is moderate; and approximately lognormal variability when it is high. In addition, high-energy gamma-ray light curves are typically analyzed in time bins of a set duration. The proposed model predicts that the fractional variability of a source will decrease proportionately to the square root of the time bin duration used in the analysis, so long as the bin duration is longer than the timescale at which individual bursts can be resolved.

\section{Autoregressive Inverse Gamma Model}\label{sec:model}

\subsection{Inverse Gamma Flux Distribution}\label{sec:derivation}

We consider a scenario in which the gamma-ray emission is dominated by discrete bursts (which might also be called flares or shots) distributed as a Poisson process. We denote the average arrival rate of the bursts as $r$ and suppose for simplicity that each burst has the same fluence $\mathcal{F}$. We consider a light curve binned in time bins of duration $\Delta T$, such that the observed flux is averaged within each time bin. Therefore, $\Delta T$ is the minimum timescale at which information about variability in the source can be determined.

We can roughly associate some characteristic timescale $t_\mathrm{shape}$ with the shape of each burst, for example, an exponential rise or decay time. We make the key assumption that $t_\mathrm{shape} \ll \Delta T$, that is, the bursts are individually unresolved on the timescale of the observations. In this scenario, all variability driven by the shapes of the individual bursts is washed out. Instead, the flux distribution derives from the Poisson statistics of the process producing the bursts.

The average number of bursts occurring in each time bin is given by

\begin{equation}\label{eq:alpha}
    \alpha = r \Delta T.
\end{equation}

As a first approximation, we assume that each bin has exactly $\alpha$ bursts. We denote the interarrival time for burst $i$ to occur as $t_{\mathrm{wait},i}$. Because the observed flux in each time bin is obtained by averaging over all of the fluxes produced by bursts in that bin, each burst will contribute a portion of the observed flux given by

\begin{equation}\label{eq:t_wait}
    F_i = \frac{\mathcal{F}}{t_{\mathrm{wait},i}}.
\end{equation}

By taking the fluence of burst $i$ to be spread out uniformly on the timescale $t_{\mathrm{wait},i}$ in its contribution to the observed flux, we have essentially further assumed that the burst shape is perfectly unresolved. We discuss the effects of modifying these assumptions in Sections~\ref{sec:discretization} and \ref{sec:discussion_limitations}.

On average, then, each burst will contribute flux $F = \mathcal{F} r$. If the binned light curve is normalized to units of flux $F_\mathrm{scale}$, where $F_\mathrm{scale}$ may be chosen arbitrarily, the average normalized flux of each burst is

\begin{equation}
    \beta_\mathrm{burst} = \frac{r\mathcal{F}}{F_\mathrm{scale}}.
\end{equation}

From Eq.~\ref{eq:t_wait}, for burst $i$, the corresponding reciprocal of the flux $F^{-1}_i$ is directly proportional to the interarrival time $t_{\mathrm{wait},i}$. In a Poisson process, the interarrival time for a single event is distributed as an exponential distribution. The same is therefore true for the reciprocal of the flux, $F^{-1} \sim \mathrm{Exp}(\beta_\mathrm{burst})$. Using the relationship between the exponential and gamma distributions, the reciprocal of the flux averaged over $\alpha$ bursts in a bin is

\begin{equation}
    F^{-1} \sim \frac{1}{\alpha} \Gamma(\alpha, \beta_\mathrm{burst}) = \Gamma(\alpha, \alpha\beta_\mathrm{burst}) =  \Gamma(\alpha, \beta),
\end{equation}

\noindent where $\Gamma(\alpha, \beta)$ is the gamma distribution with shape parameter $\alpha$ and rate parameter $\beta$, where

\begin{equation}\label{eq:beta}
   \beta = \alpha \beta_\mathrm{burst} = \frac{\alpha r\mathcal{F}}{F_\mathrm{scale}}. 
\end{equation}

Since $F = (F^{-1})^{-1}$, it follows from the relationship between the gamma and inverse gamma distributions that

\begin{equation}
    F \sim \mathrm{InvGamma}(\alpha, \beta),
\end{equation}

\noindent where $\mathrm{InvGamma}(\alpha, \beta)$ is the inverse gamma distribution with shape parameter $\alpha$ and scale parameter $\beta$.

From Eqs.~\ref{eq:alpha} and \ref{eq:beta}, the parameters of the inverse gamma distribution are related to the physical characteristics of the bursts as

\begin{equation}\label{eq:physical_parameters}
\begin{split}
    r &= \frac{\alpha}{\Delta T}\\
    \mathcal{F} &= \frac{\beta}{\alpha^2} F_\mathrm{scale} \Delta T.
\end{split}
\end{equation}

\subsection{Incorporating Autoregression}

To model a stationary time series possessing AR(1) autoregressive structure and an inverse gamma flux distribution, we make use of the existing literature on conditional linear AR(1) processes \citep{Grunwald2000}, as the simplest formulation that can produce a non-Gaussian time series with autoregressive structure. We consider the time series of $F^{-1}$, allowing us to take advantage of existing models of gamma-distributed AR(1) processes. The flux light curve, with an inverse gamma flux distribution, can then be obtained by applying a reciprocal transformation to the time series.

In the standard AR(1) process, a normal innovation produces a normal marginal distribution, but the gamma distribution does not have this property. A more complex model is required to produce an AR(1) process with a gamma marginal distribution. We model the autoregressive structure of the light curve using the process proposed by \citet{Sim1990}, in which the time dependence has the form

\begin{equation}\label{eq:sim1990}
\begin{split}
    &F_i^{-1} = \Gamma(N(F_{i - 1}^{-1}) + \alpha, \beta),\\
    &N(F^{-1}) \sim \mathrm{Pois}(\phi \beta F^{-1}).
\end{split}
\end{equation}

In this approach, known as thinning, rather than multiplying $F_{i - 1}^{-1}$ by a constant $\phi$, it is instead reduced using a stochastic function taking $\phi$ as a parameter. The process is time-reversible. As shown by \citet{Sim1990}, the autocorrelation function is $\mathrm{Corr}(F_{i + j}^{-1}, F_i^{-1}) = \phi^j$ and the resulting marginal distribution is

\begin{equation}
    f(F^{-1}) = \Gamma\left(\alpha, (1 - \phi)\beta\right).
\end{equation}

The flux distribution is therefore

\begin{equation}
    f(F) = \mathrm{InvGamma}(\alpha, (1 - \phi)\beta).
\end{equation}

The autocorrelation parameter $\phi$ can be used to estimate an autocorrelation timescale $\tau$,

\begin{equation}
    \phi = e^{-\Delta T/\tau},
\end{equation}

\noindent or as a function of $\phi$,

\begin{equation}
    \tau = \frac{\Delta T}{\ln{(1/\phi)}}.
\end{equation}

Several other models for a gamma AR(1) process have been proposed, possessing different statistical properties. In the GAR(1) model \citep{Gaver1980, Lawrance1982, Walker2000}, a marginal gamma distribution is obtained by considering an innovation constructed to represent a shot-noise process with an exponential distribution of shot amplitudes. The process is not time-reversible and exhibits peaks followed by geometrically decaying runs of values. While we do not consider this model well suited to describe fluxes averaged over time bins containing multiple unresolved bursts, it may have useful applications in other astrophysical contexts in which fast rising and exponentially decaying flares are individually resolved.

In the BGAR(1) process \citep{Lewis1989}, the innovation has a gamma distribution, and a gamma marginal distribution is obtained by replacing the constant autoregression coefficient $\phi$ with a random variable following a beta distribution with expected value $\phi$. The BGAR(1) process is time-reversible. However, the marginal distribution resulting from this process is independent of $\phi$, which is inconsistent with the dependence on $\phi$ in the marginal distribution that might be expected by considering the limiting Gaussian case (Eq.~\ref{eq:gaussian_ar1}). Here, the autoregressive structure could be thought of as making it likely that nearby time steps sample similar values of the innovation PDF, which is difficult to interpret physically.

\subsection{Accounting for the Discretization of Bursts in Time Bins}\label{sec:discretization}

As we have described the model so far, each time bin contains exactly the average number of bursts $\alpha$. More realistically, the number of bursts would vary from time bin to time bin following a Poisson distribution with parameter $\alpha$. This scenario has two main implications. First, a time bin could contain no bursts, so it would have no associated flux. To allow this, we can describe the inverse flux using a mixed distribution, with finite probability mass $p_0 = e^{-\alpha}$ at $F^{-1} = 0$ and continuous probability density integrating to $p_\Gamma = 1 - e^{-\alpha}$ for $F^{-1} > 0$.

\begin{figure*}[htb]
    \centering
    \includegraphics[width=0.8\textwidth]{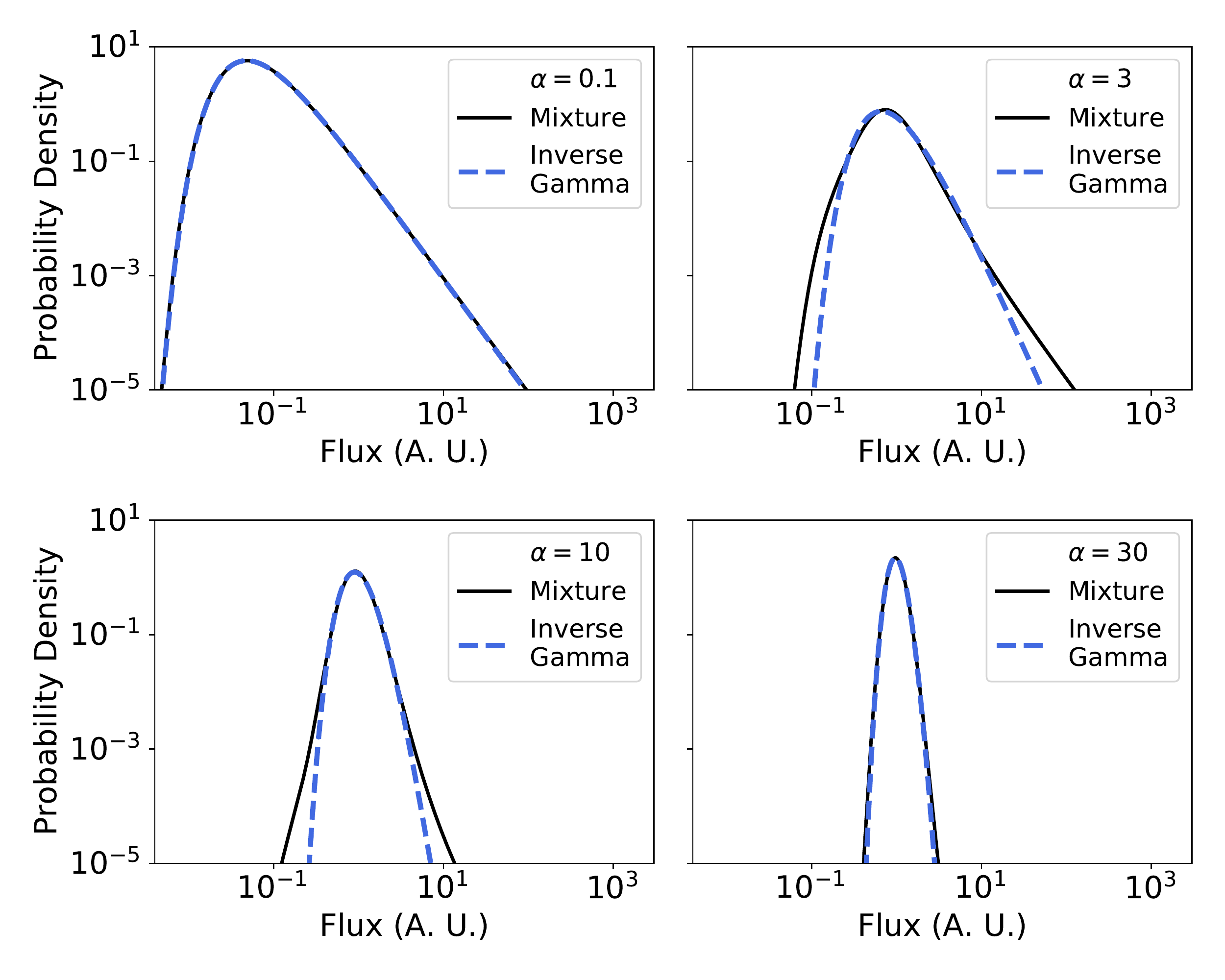}
    \caption{Best-fit inverse gamma flux distributions to Poisson-weighted mixtures of inverse gamma distributions, using the approximation method described in Sec.~\ref{sec:discretization}.}
    \label{fig:mixture_gamma_approximation}
\end{figure*}

Second, for bins that do have bursts, because the number of bursts is different in each bin, the continuous part of the inverse-flux distribution is no longer a single gamma distribution with shape parameter $\alpha$, but a weighted mixture of gamma distributions with shape parameters k = 1, 2, 3, etc. While the flux distribution resulting from this mixture is not exactly inverse gamma in general, as shown in Figure~\ref{fig:mixture_gamma_approximation}, it can be approximated well by a single inverse gamma distribution.

The properties of the resulting process are derived in Appendix~\ref{appendix:derivations}. The distribution of $F^{-1} > 0$ is given by a gamma distribution with shape parameter

\begin{equation}
    \alpha_\mathrm{obs} = \frac{A(\alpha)}{1 - e^{-\alpha}}\alpha,
\end{equation}

\noindent and rate parameter

\begin{equation}
    \beta_\mathrm{obs} = A(\alpha)(1 - \phi)\beta,
\end{equation}

\noindent where

\begin{equation}
    A(\alpha) = \left( \frac{\alpha}{(1 - e^{-\alpha})^2} \sum_{l = 1}^{\infty} \frac{\alpha^l e^{-\alpha}}{l!} \frac{1}{l} \right)^{-1}.
\end{equation}

\begin{figure}[ht]
    \centering
    \includegraphics[width=0.45\textwidth]{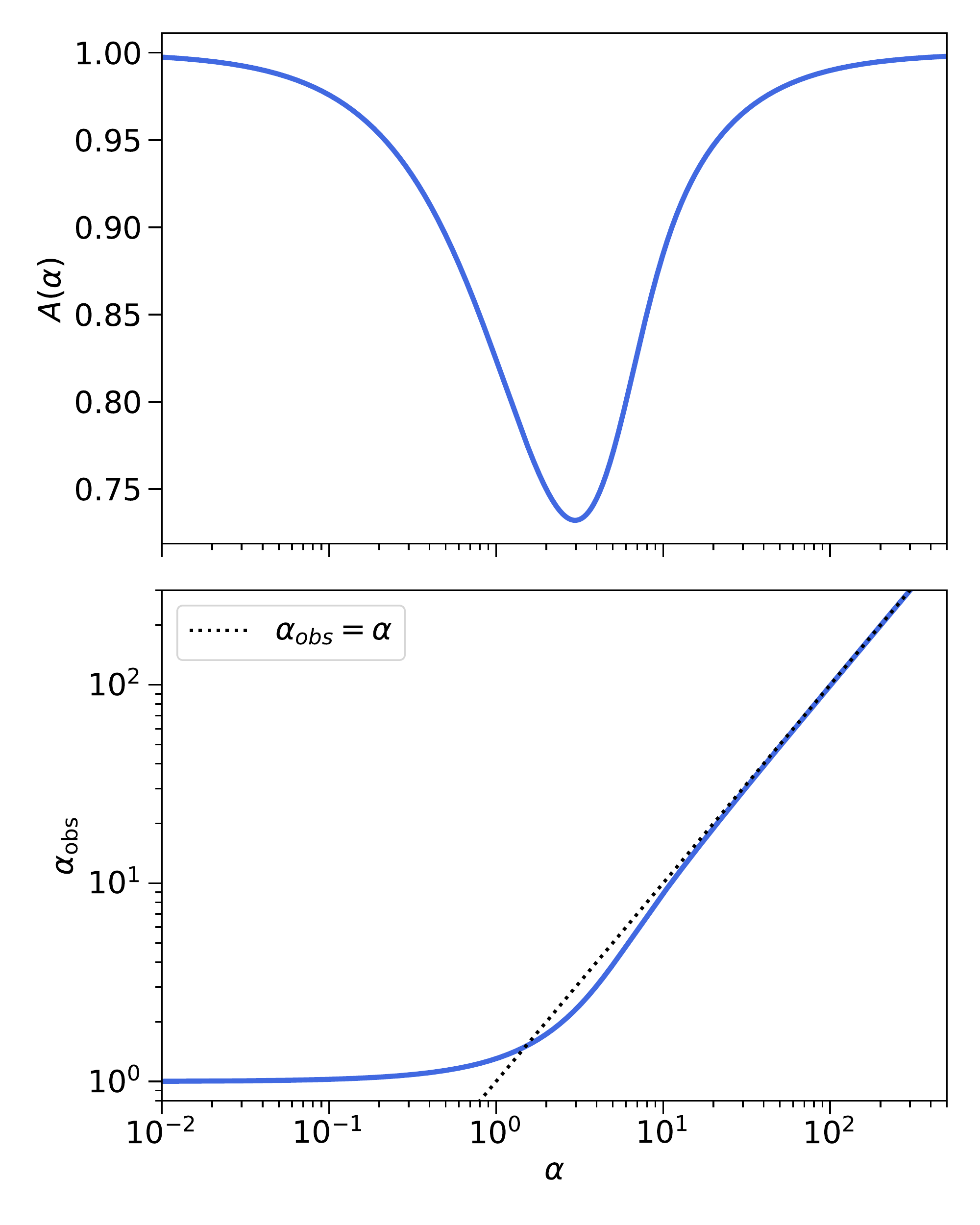}
    \caption{\textit{Top}: Modification factor $A(\alpha)$ of the inverse gamma parameters to account for a Poisson distribution of bursts, as a function of the intrinsic average burst rate $\alpha$. \textit{Bottom}: The apparent burst rate $\alpha_\mathrm{obs}$ as a function of $\alpha$.}
    \label{fig:a_alpha_obs}
\end{figure}

In Figure~\ref{fig:a_alpha_obs}, $A(\alpha)$ and $\alpha_\mathrm{obs}$ are plotted as a function of $\alpha$. As shown in the figure, $A(\alpha) \lesssim 1$ for all $\alpha$, reaching a minimum of $A(\alpha) \approx 0.73$  at $\alpha \approx 2.98$, and approaching 1 in the limits $\alpha \to 0$ and $\alpha \to \infty$. As a result, $\alpha_\mathrm{obs}$ approaches 1 as $\alpha \to 0$ and approaches $\alpha$ as $\alpha \to \infty$.

For time bins containing bursts, the behavior of the time series is described by Eq.~\ref{eq:sim1990} with a simple change of parameters. Since $\alpha_\mathrm{obs}$ and $\beta_\mathrm{obs}$ are the parameters of the marginal distribution, the process parameters are $\alpha_\mathrm{obs}$ and $\beta_\mathrm{obs}/(1 - \phi) = A(\alpha)\beta$, giving

\begin{equation}\label{eq:sim1990_modified}
\begin{split}
    &F_i^{-1} = \Gamma(N(F_{i - 1}^{-1}) + \alpha_\mathrm{obs}, A(\alpha)\beta),\\
    &N(F^{-1}) \sim \mathrm{Pois}(\phi A(\alpha)\beta F^{-1}).
\end{split}
\end{equation}

To accommodate bins with $F^{-1} = 0$ while preserving the AR(1) autocorrelation structure, the modified process can be described as a Markov process with two states, $F^{-1} = 0$ for bins without bursts and $F^{-1} \sim \Gamma(\alpha_\mathrm{obs}, \beta_\mathrm{obs})$ for bins with bursts. The state transition probabilities $s_i \in \{0, \Gamma\} \to s_{i + 1} \in \{0, \Gamma\}$ are given by

\begin{equation}\label{eq:transition_probs}
\begin{split}
    p_{00} &= 1 - p(\phi, \alpha)(1 - e^{-\alpha})\\
    p_{0\Gamma} &= p(\phi, \alpha)(1 - e^{-\alpha})\\
    p_{\Gamma0} &= p(\phi, \alpha)e^{-\alpha}\\
    p_{\Gamma\Gamma} &= 1 - p(\phi, \alpha)e^{-\alpha},
\end{split}
\end{equation}

\noindent where

\begin{equation}
    p(\phi, \alpha) =  \frac{1 - \phi}{1 + \frac{\phi}{\alpha}A^{-1}(\alpha)(1 - e^{-\alpha})}.
\end{equation}


\subsection{``Loner Flares'' at Low Burst Rates}

When the burst rate is very low, most time bins contain no bursts, and the light curve consists of isolated flares separated by intervals of no emission. The model parameters $\alpha$, $\beta$, and $\phi$ can be extracted from the properties of the isolated flares by considering the limiting case where $\alpha \ll 1$. Specifically, to first order in $\alpha$, the parameters of the observed distribution become

\begin{subequations}
\begin{align}
    \alpha_\mathrm{obs} &\approx 1 + \frac{\alpha}{4},\\
    \beta_\mathrm{obs} &\approx \left(1 - \frac{\alpha}{4} \right)(1 - \phi)\beta,\label{eq:lowalpha_beta}\\
    p_\Gamma &\approx \alpha,\label{eq:lowalpha_pgamma}\\
    p(\phi, \alpha) &\approx \frac{1 - \phi}{1 + \phi}\left(1 + \frac{\phi}{1 + \phi}\frac{\alpha}{4} \right).
\end{align}
\end{subequations}

The expected fraction of time bins that contain flaring gamma-ray emission equals $p_\Gamma$ and so provides a direct estimate of $\alpha$ from Eq.~\ref{eq:lowalpha_pgamma}. The average length of flares in units of $\Delta T$ is given by

\begin{equation}
    N_\mathrm{avg} = \frac{1}{p_{\Gamma 0}} \approx \frac{1 + \phi}{1 - \phi}\left(1 + \frac{\alpha}{1 + \phi} \right).
\end{equation}

In particular, if $\alpha$ is very small, $\phi$ can be roughly estimated as

\begin{equation}
    \phi \approx \frac{N_\mathrm{avg} - 1}{N_\mathrm{avg} + 1}.
\end{equation}

A measurement of $\beta_\mathrm{obs}$, thereby giving an estimate of $\beta$ from Eq.~\ref{eq:lowalpha_beta}, can then be obtained by fitting an inverse gamma distribution to the observed flaring flux distribution.

\subsection{Lognormal Behavior at High Burst Rates}\label{sec:lognormal_approximation}

When the burst rate is very high, the law of large numbers applies: every time bin contains a number of bursts close to the expected value, and the dynamic range of variability is reduced. The inverse flux distribution approaches a normal distribution, resulting in a flux distribution that is approximately lognormal. Specifically, if $\alpha$ is large, the parameters of the observed distribution become $\alpha_\mathrm{obs} \approx \alpha$, $\beta_\mathrm{obs} \approx (1 - \phi)\beta$, and $p_\Gamma \approx 1$, and

\begin{equation}
    \Gamma(\alpha_\mathrm{obs}, \beta_\mathrm{obs}) \to \mathcal{N}\left(\frac{\alpha_\mathrm{obs}}{\beta_\mathrm{obs}}, \frac{\alpha_\mathrm{obs}}{\beta_\mathrm{obs}^2}\right),
\end{equation}

\noindent where $\mathcal{N}(\mu, \sigma^2)$ is the normal distribution with mean $\mu$ and variance $\sigma^2$. Then, to a level of $s$ standard deviations, the flux $F$ is constrained such that

\begin{equation}
    \left| \frac{1}{F} - \frac{\alpha_\mathrm{obs}}{\beta_\mathrm{obs}} \right| \lesssim \frac{s \sqrt{\alpha_\mathrm{obs}}}{\beta_\mathrm{obs}},
\end{equation}

\noindent from which we obtain

\begin{equation}
    \left| \frac{\beta_\mathrm{obs}}{\alpha_\mathrm{obs}}\frac{1}{F} - 1 \right| \lesssim \frac{s}{\sqrt{\alpha}}.
\end{equation}

Letting $\Delta \equiv \beta_\mathrm{obs}/(\alpha_\mathrm{obs}F) - 1$, then within a significance level of $\sim5$ standard deviations, $\left| \Delta \right| \lesssim 1$ when $\alpha \gtrsim 25$.

For a random variable $X$ with PDF $f(X)$, the reciprocal variable $Y = 1/X$ has the PDF $g(Y) = Y^{-2}f(1/Y)$. Since the inverse flux is normally distributed, the flux $F$ has the PDF

\begin{equation}
\begin{split}
    F &\sim \frac{1}{F^2} \frac{1}{(\sqrt{\alpha_\mathrm{obs}}/\beta_\mathrm{obs})\sqrt{2\pi}} e^{-\frac{1}{2}\left(\frac{(1/F) - (\alpha_\mathrm{obs}/\beta_\mathrm{obs})}{\sqrt{\alpha_\mathrm{obs}}/\beta_\mathrm{obs}}\right)^2}\\
    &= (1 + \Delta)\frac{1}{F}\sqrt{\frac{\alpha_\mathrm{obs}}{2\pi}} e^{-\frac{1}{2}\alpha_\mathrm{obs}\Delta^2}\\
    &\approx \frac{1}{F}\sqrt{\frac{\alpha_\mathrm{obs}}{2\pi}}  e^{-\frac{1}{2}\alpha_\mathrm{obs}\ln^2{(1 + \Delta)}}\\
    &= \frac{1}{F}\sqrt{\frac{\alpha_\mathrm{obs}}{2\pi}}  e^{-\frac{1}{2}\alpha_\mathrm{obs}(\ln{F} - \ln{(\beta_\mathrm{obs}/\alpha_\mathrm{obs}}))^2}\\
    &= \mathrm{Lognormal}(\mu, \sigma^2),
\end{split}
\end{equation}

\noindent where

\begin{equation}\label{eq:lognormal_approximation}
\begin{split}
    \mu = \ln{\frac{\beta_\mathrm{obs}}{\alpha_\mathrm{obs}}} &\approx \ln{\frac{(1 - \phi)\beta}{\alpha}},\\
    \sigma^2 = \frac{1}{\alpha_\mathrm{obs}} &\approx \frac{1}{\alpha}.
\end{split}
\end{equation}

To first order, the flux distribution can be approximated as a lognormal distribution, the parameters of which can be interpreted in terms of the underlying burst process. Using Eq.~\ref{eq:physical_parameters}, the burst parameters can be estimated from the lognormal fit as

\begin{equation}
\begin{split}
    r &\approx \left(\sigma^2 \Delta T \right)^{-1}\\
    \mathcal{F} &\approx \frac{1}{1 - \phi} \sigma^2 e^\mu F_\mathrm{scale} \Delta T.
\end{split}
\end{equation}

\subsection{Fractional Variability}

The shape of the flux distribution depends on the shape parameter $\alpha$. From Eq.~\ref{eq:alpha}, $\alpha$ is the product of two factors: the burst rate $r$, which is an intrinsic physical parameter, and the time bin duration $\Delta T$, which is an extrinsic choice made in the analysis. This fact enables us to predict how the fractional variability of a given source should scale when analyzed using time bins of different durations.

The mean of an inverse gamma distribution is

\begin{equation}
    \mu = \frac{\beta}{\alpha - 1},
\end{equation}

\noindent and its variance is

\begin{equation}
    \sigma^2 = \frac{\beta^2}{(\alpha - 1)^2(\alpha - 2)}.
\end{equation}

The mean is defined only for $\alpha > 1$ and the variance for $\alpha > 2$. Then, for $\alpha > 2$, the expected fractional variability is

\begin{equation}
\begin{split}
    F_\mathrm{var} &= \frac{\sigma}{\mu} = (\alpha_\mathrm{obs} - 2)^{-1/2} \approx (\alpha - 2)^{-1/2}\\
    &= r^{-1/2}\left(1 - \frac{2}{r\Delta T} \right)^{-1/2} \Delta T^{-1/2}.
\end{split}
\end{equation}

The fractional variability of a given source should scale with the time binning as $F_\mathrm{var} \propto \Delta T^{-1/2}$, at least in the regime $\Delta T \gtrsim \max(2/r, t_\mathrm{shape})$.

\section{Representative Light Curves}\label{sec:simulation}

\begin{figure*}[htp]
    \centering
    \includegraphics[width=0.49\textwidth]{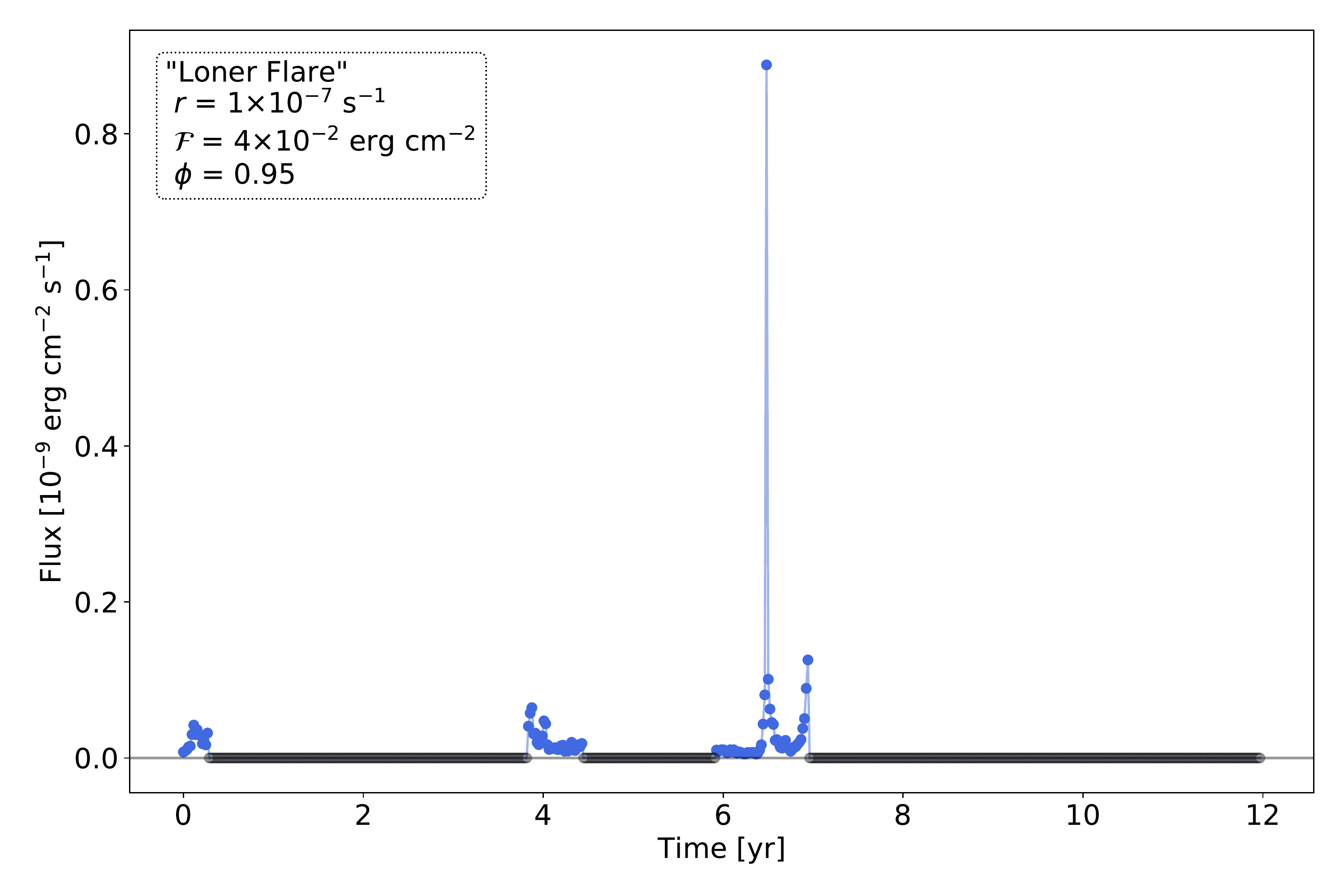}
    \includegraphics[width=0.49\textwidth]{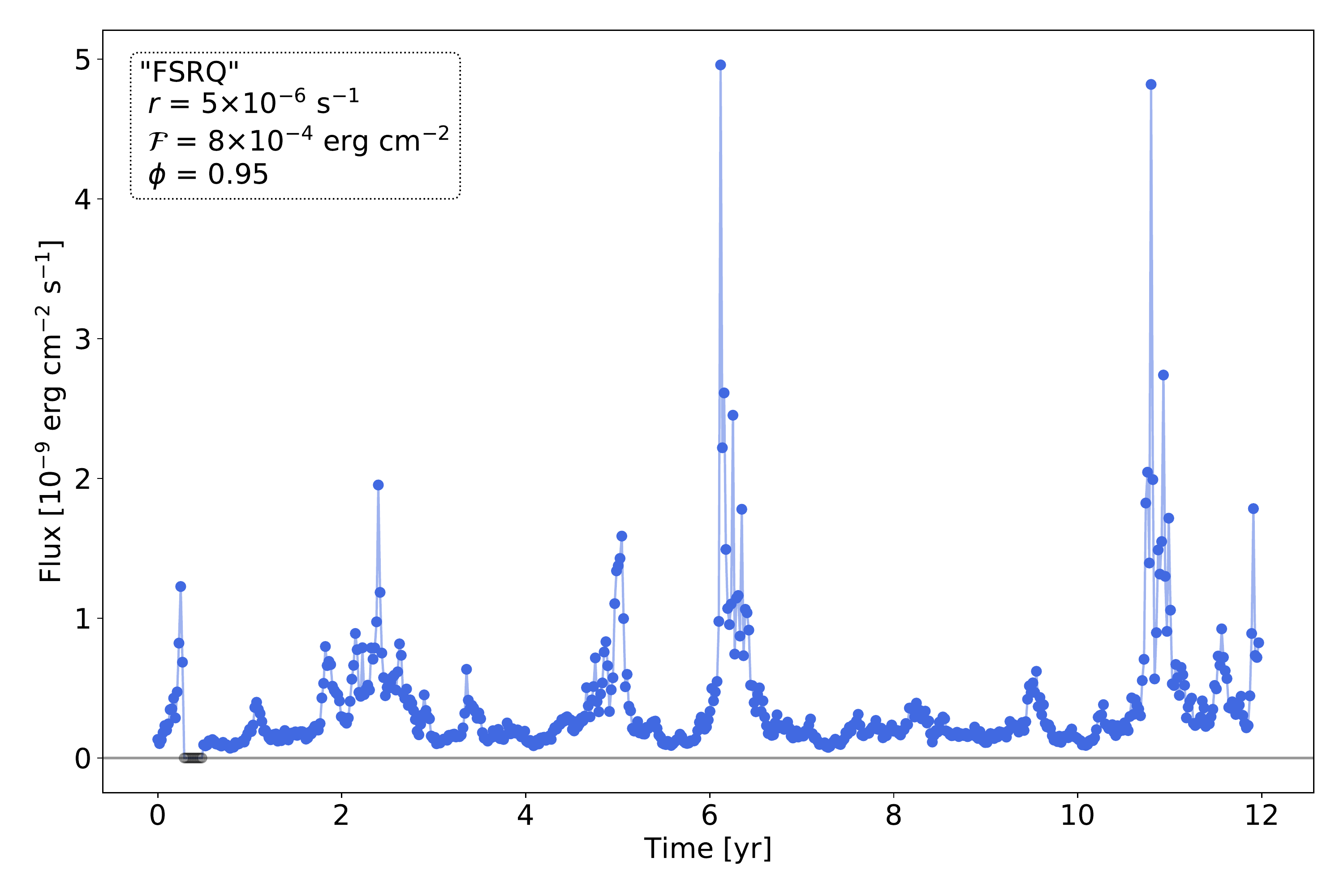}
    \includegraphics[width=0.49\textwidth]{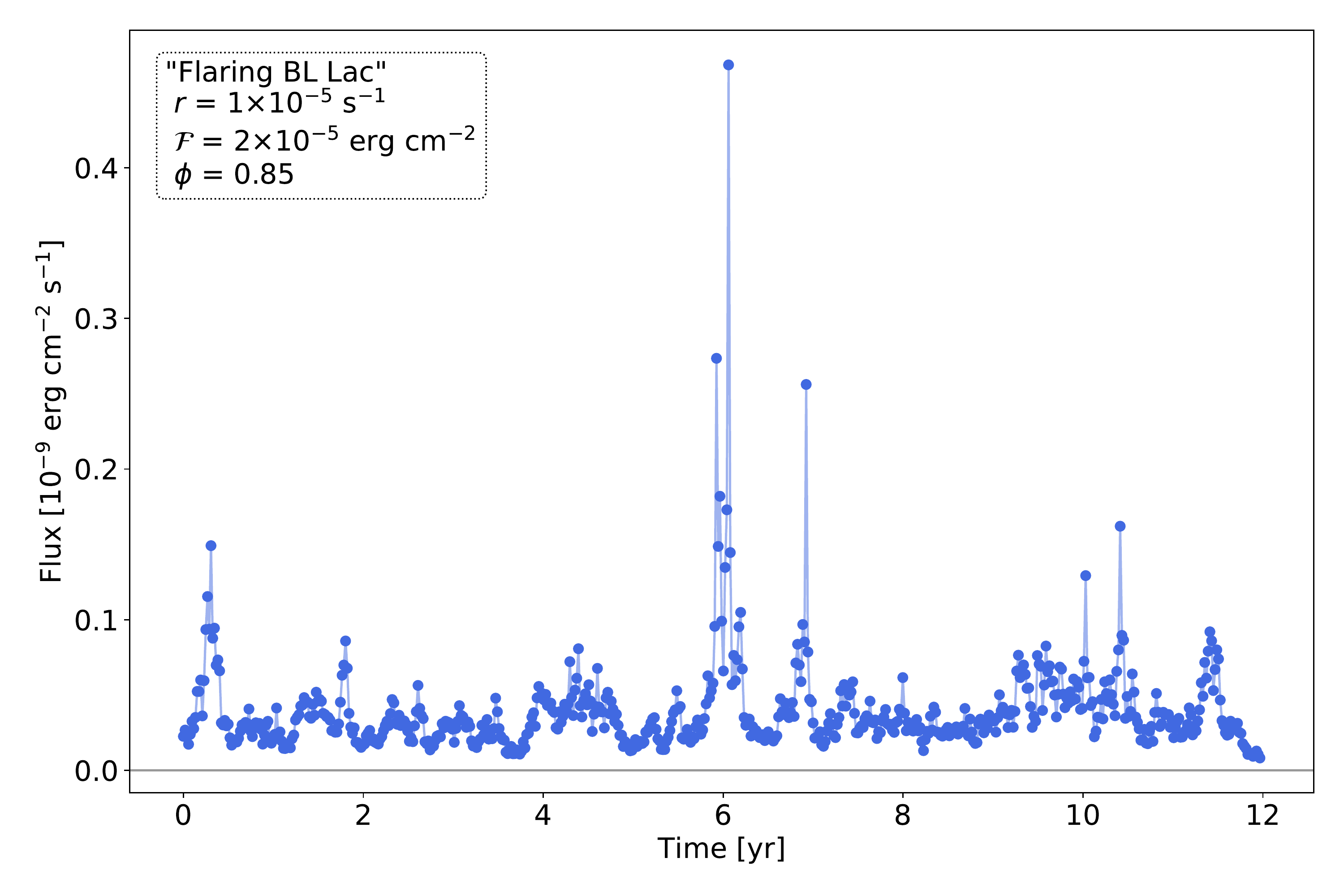}
    \includegraphics[width=0.49\textwidth]{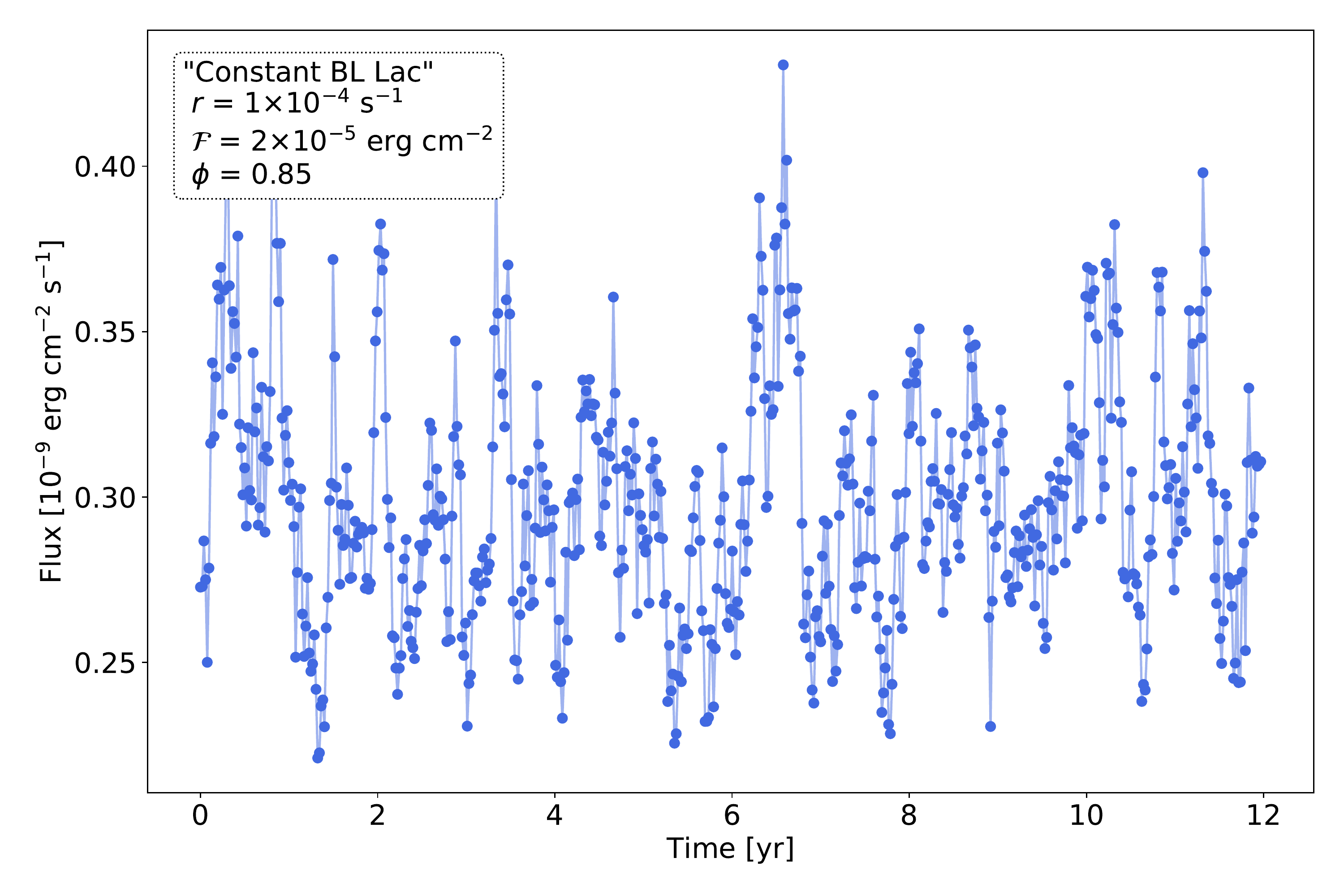}
    \caption{\textit{Top Left}: Simulated light curve meant to resemble a ``loner flare'' blazar. \textit{Top Right}: Simulated light curve meant to resemble an FSRQ, with parameters chosen based on gamma-ray observations of 3C~279. \textit{Bottom Left}: Simulated light curve meant to resemble a ``flaring BL Lac'' object, such as 1ES~1215+303. \textit{Bottom Right}: Simulated light curve meant to resemble a ``constant BL Lac'' object.}
    \label{fig:lightcurves}
\end{figure*}

\begin{figure*}[htp]
    \centering
    \includegraphics[width=0.9\textwidth]{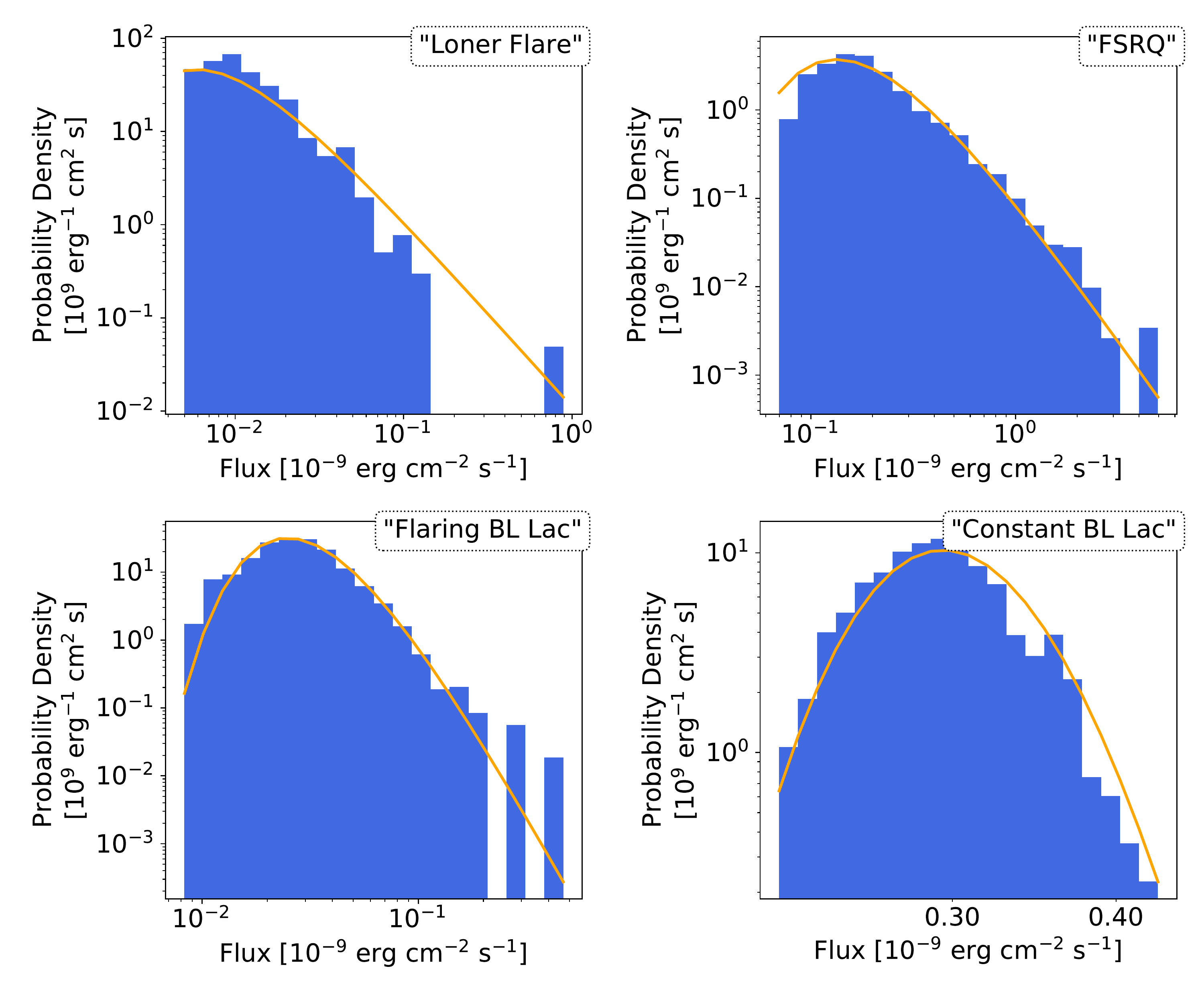}
    \caption{Flux distributions corresponding to the four simulated light curves. The expected PDF is shown as an orange line in each plot.}
    \label{fig:flux_distributions}
\end{figure*}

\begin{figure*}[htp]
    \centering
    \includegraphics[width=0.9\textwidth]{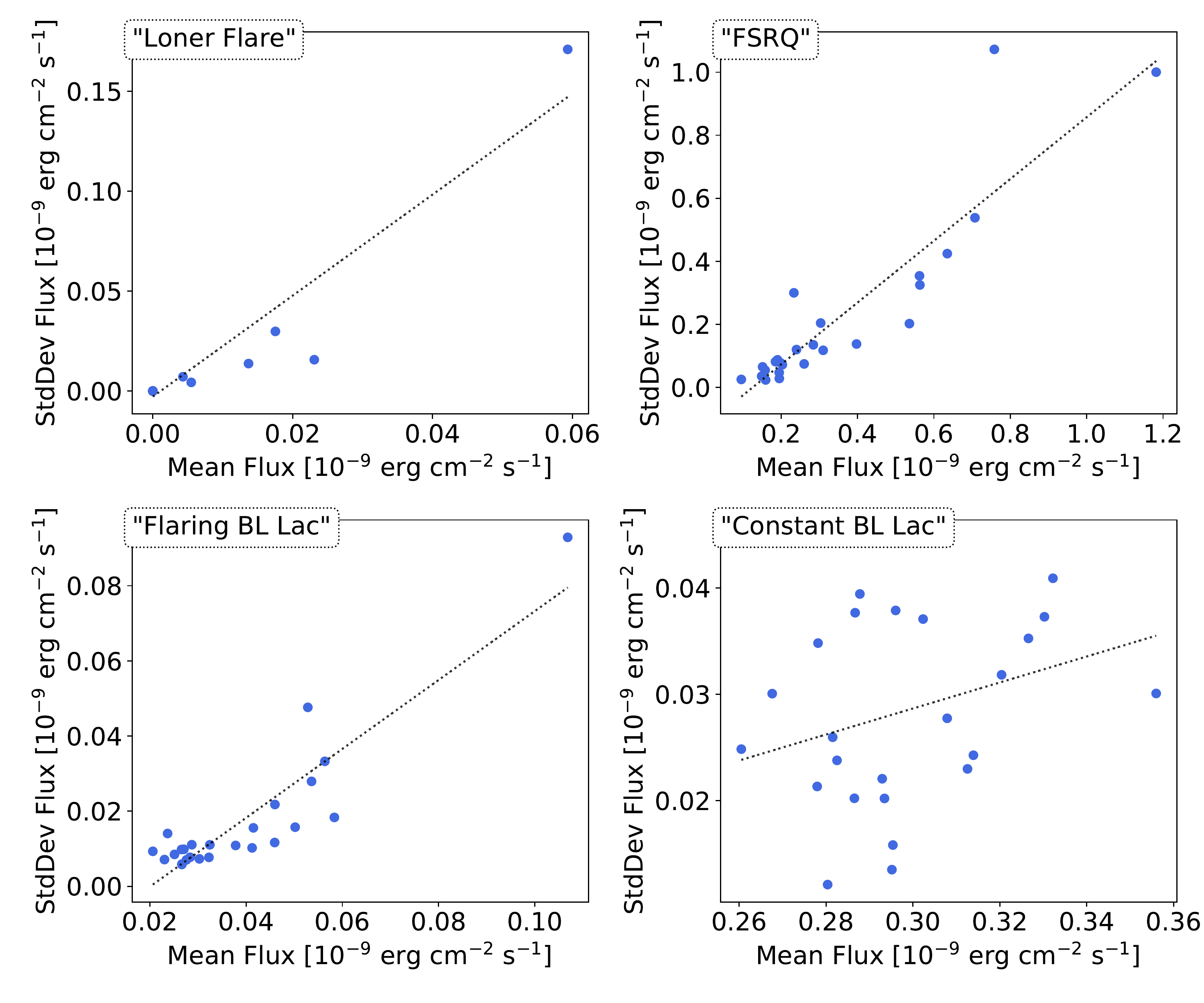}
    \caption{RMS-flux relations corresponding to the four simulated light curves. A linear least-squares fit is shown with a dotted black line.}
    \label{fig:rms_flux}
\end{figure*}

To illustrate the process, we show simulated 12-year weekly light curves representing four different source types in Figure~\ref{fig:lightcurves}. The light curves were generated following the formulas given in Eqs.~\ref{eq:sim1990_modified} and \ref{eq:transition_probs}. Each light curve represents only one possible realization of the corresponding stochastic process.

To generate a representative light curve for an FSRQ, we estimated $r$ and $\mathcal{F}$ from the observations reported by \citet{Adams2022} of flares of 3C~279, giving $r \approx 24/ 54~\mathrm{bursts}~\mathrm{day}^{-1} \approx 5\times10^{-6}~\mathrm{s}^{-1}$ and $\mathcal{F} \approx 0.85\times 10^{-3}~\mathrm{erg}~\mathrm{cm}^{-2}$. We set a fairly high autocorrelation parameter of $\phi = 0.95$, consistent with the absence of a resolvable spectral break in the gamma-ray PSD of 3C~279 as determined both by CARMA modeling \citep{Ryan2019} and by the power spectral response method \citep{Goyal2021}. With these parameters, $\alpha_\mathrm{obs} \approx 2.3$, similar to the best-fit values of $\lambda + 1 = 1.7$ and 2.0 reported respectively by \citet{Tavecchio2020} and \citet{Adams2022}.

The resulting simulated light curve is shown in the top right of Figure~\ref{fig:lightcurves}. The pattern of variability, showing several flaring periods separated by intervals of relatively quiescent activity, indeed appears to resemble an actual light curve of an FSRQ like 3C~279. The overall normalization of the energy flux during the flares and quiescent periods is also similar to that seen in the actual light curve of that source.

Figure~\ref{fig:lightcurves}, top left, illustrates a source exhibiting even more extreme variability, with isolated flares separated by intervals of complete quiescence. For this simulation, the burst rate was decreased by a factor of 50 and the burst fluence increased by the same. This light curve resembles that of a so-called ``loner flare'' blazar characterized by the detection of a single high-flux flaring event in the long-term gamma-ray light curve \citep{Wang2020}.


The autoregressive inverse gamma model may also be able to reproduce the less extreme variability associated with BL Lac objects. For example, an extensive analysis of the multiwavelength variability of the flaring high-synchrotron-peaked BL Lac object 1ES~1215+303 was performed by \citet{Valverde2020}, who found that its flux distribution was described better by a lognormal distribution than by a normal distribution. They obtained a lognormal distribution with a best-fit value of $\sigma = 0.43 \pm 0.02$, which, using Eq.~\ref{eq:lognormal_approximation}, can be reproduced by an inverse gamma distribution with $\alpha \approx 5.4$.

With this motivation, we generated simulated light curves with parameters chosen to represent two ``BL Lac'' objects, a ``flaring'' source with $\alpha_\mathrm{obs} \approx 5$ intended to be reminiscent of a source like 1ES~1215+303, and a ``constant'' source with $\alpha_\mathrm{obs} \approx 60$. These simulated light curves are shown in Figure~\ref{fig:lightcurves}, bottom left and right, respectively. Again, the simulated light curves appear to resemble real light curves.

The flux distributions corresponding to the simulated light curves shown in Figure~\ref{fig:lightcurves} are plotted in Figure~\ref{fig:flux_distributions}, along with the expected PDF for each.

We examined the RMS-flux relation for each of the simulated light curves. The light curves were binned in intervals of 25 data points and the mean and standard deviation was calculated within each interval. The standard deviation for each interval is plotted against the mean in Figure~\ref{fig:rms_flux}. An approximately linear RMS-flux relation can be observed in each case. The ``constant BL Lac'' light curve, with the least skewed flux distribution, exhibits the RMS-flux relation with the most scatter.

\section{Discussion}\label{sec:discussion}

\subsection{Applicability and Extensions of the Model}\label{sec:discussion_limitations}

The autoregressive inverse gamma model for blazar variability presented in this work is based on several simplifying assumptions which should be considered when interpreting the model and the resulting flux distribution. An in-depth exploration of the effects of modifying these assumptions is beyond the scope of the simple statistical model presented in this work, and may be better addressed in the context of detailed physical simulations.

\begin{figure*}[htp]
    \centering
    \includegraphics[width=0.9\textwidth]{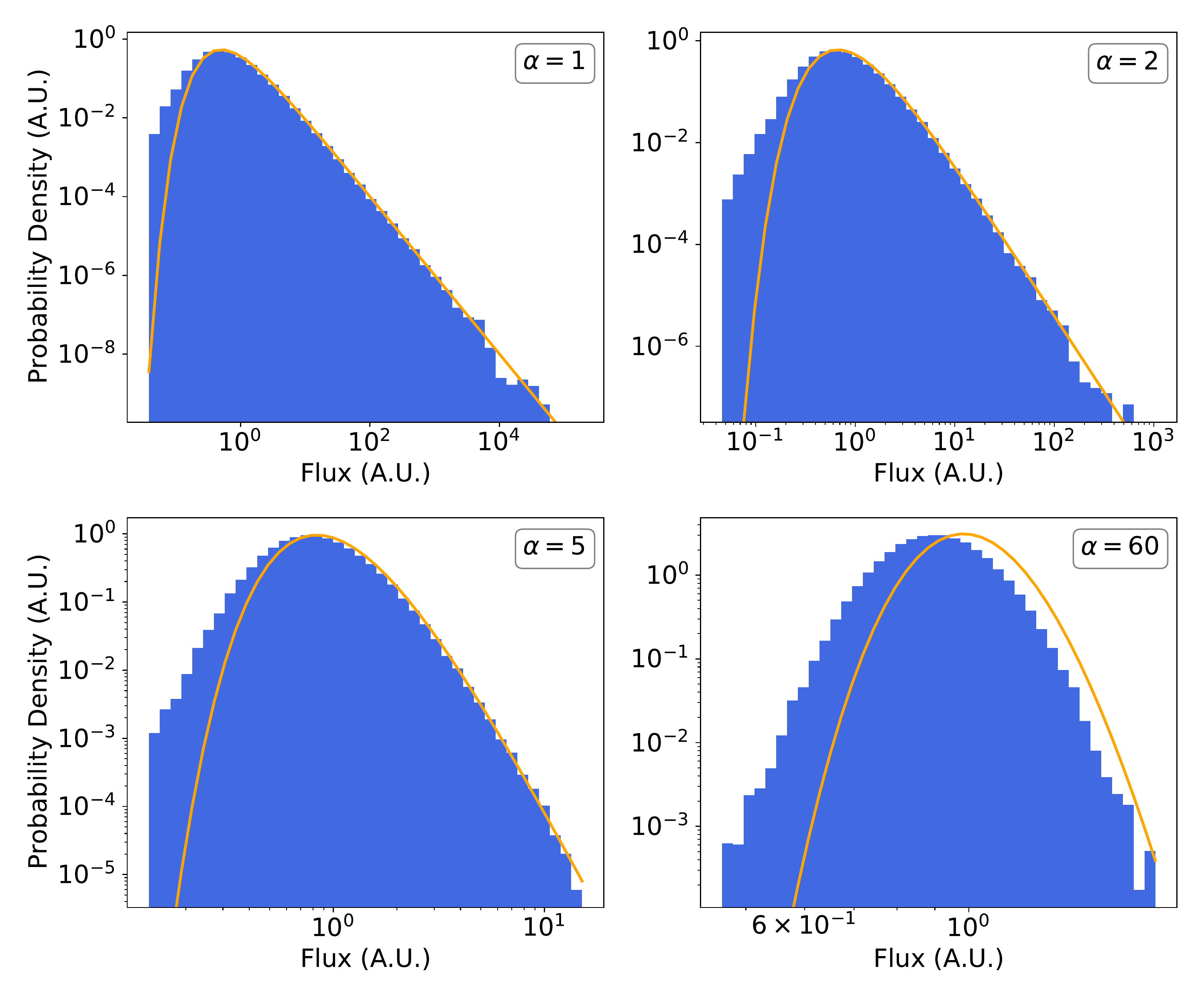}
    \caption{Simulated flux distributions incorporating variations in burst fluence of up to an order of magnitude. The corresponding unmodified inverse gamma distributions are shown with orange lines.}
    \label{fig:fluence_variability}
\end{figure*}

First, we have assumed that all bursts have exactly the same fluence.
More realistically, the fluence may vary from burst to burst. An illustration of how varying the fluence can affect the flux distribution is shown in Figure~\ref{fig:fluence_variability}. Distributions are shown for $\alpha = 1$, 2, 5 and 60, similar to the simulated sources considered in Section~\ref{sec:simulation}, scaled such that $\beta = \alpha$. We numerically simulated a modification of the basic scenario described in Sec.~\ref{sec:derivation}, without autoregression, for variable burst fluence by multiplying the mean fluence $\mathcal{F}$ by a random scale factor for each burst. The scale factors were drawn from a truncated Gaussian distribution with $\mu=1$ and $\sigma=0.275$. The distribution was truncated at $-3 \sigma$ to avoid negative fluctuations and at $+3 \sigma$ to preserve symmetry. We chose a value of $\sigma=0.275$ to yield a fluence distribution varying by one order of magnitude.

Allowing the fluence to vary produces an excess at the low-flux end of the flux distribution, causing it to cut off less sharply, while preserving the heavy high-flux tail. This occurs because the fluence fluctuations are dominated by the burst arrival-time fluctuations except at the lowest fluxes, at least for small $\alpha$. For the much narrower distribution with $\alpha=60$, the entire distribution is noticeably shifted, while retaining an approximately lognormal shape.

An additional low-flux component could also arise in the flux distribution for other reasons. Such a component might come about, for example, through the flux contributed by a subdominant population of weak bursts with low fluence. Also, if the bursts have relatively long rise or decay times, ``leakage'' of the flux from one bin to the next could result in a similar effect. The effect of such a component would be most apparent as an excess at the low end of the flux distribution, in addition to potentially adding a small flux component to the time bins that are considered here to have no flux. However, an additional low-flux component may be difficult to identify in data, as it must be disentangled from potential artifactual contributions, such as statistical fluctuations or imperfectly modeled gamma-ray flux coming from nearby sources in the field of view.

In addition, as shown in Figure~\ref{fig:mixture_gamma_approximation}, the exact Poisson-weighted mixture distribution has in general heavier tails than the best-fit inverse gamma distribution. For this reason, a pure inverse gamma flux distribution may somewhat underestimate the true expected variability.

We have also assumed that the burst arrival times follow a Poisson process, but in reality there may be physical constraints on how often bursts can occur. For example, if a lower bound exists on the time that must elapse between two bursts, that would place an upper bound on the imputed flux of a single burst as well as on the number of bursts that can occur in a time bin, inducing a cut-off at the high-flux end of the flux distribution.

The high-flux end of the flux distribution is critical for distinguishing the inverse gamma distribution from other proposed distributions, such as the lognormal. However, high-flux time bins are rare. The longer the light curve, the more effectively the high-flux part of the flux distribution can be studied. Some previous studies that compared normal and lognormal distributions have excluded flaring intervals from the flux distribution so as not to overly favor the heavier-tailed lognormal distribution \citep[e.g.][]{Giebels2009, Valverde2020}. However, when only considering heavy-tailed distributions, these intervals should not be excluded.

The derivation of the statistical model presented here requires that individual bursts are individually unresolved. For this reason, the model is only applicable at time scales longer than the putative burst shape timescale $t_\mathrm{shape}$. For very bright sources, it may be possible to estimate $t_\mathrm{shape}$ from an analysis of short-timescale variability during high-amplitude flares, but for weaker sources, this quantity may be more difficult to measure. At timescales shorter than $t_\mathrm{shape}$, the observed variability would depend on the characteristics of the bursts as well as on the statistics of their stochastic production process.

For example, in the study done by \citet{Adams2022}, a time binning of 1 day was used when studying the flux distribution of 3C~279, which is shorter than the longest timescale of $\sim$100 hr found by fitting exponential profiles to individual flare components. In this case, care must be taken when interpreting the inverse gamma fit parameters using the model presented in this work.

In this paper, we exclusively considered AR(1) processes. Some gamma-ray blazars may exhibit possible year-scale \mbox{(quasi-)periodicities} \citep[e.g.][]{Penil2020, Rueda2022} or trends \citep[e.g.][]{Valverde2020}. Future work could investigate the possibility of extending the model by adding deterministic trend or periodic terms, higher-order autoregressive terms, or moving average terms. Sources with possible year-scale quasi-periodic oscillations could be modeled using a second-order autoregressive process. Alternatively, one could study nonlinear time-series structures in which the innovation depends on the flux, such as in the SDE proposed by \citet{Tavecchio2020} or the autoregressive gamma process (ARG) of \citet{Gourieroux2006}.

\subsection{Connection to Theoretical Models}

The simulations shown in Section~\ref{sec:simulation} suggest that a single stochastic process can describe both flares and quiescent periods in light curves, consistent with the previous studies that have modeled blazar light curves using autoregressive processes \citep[e.g.][]{Ryan2019}. The process proposed in this work has three free parameters, connected to the average burst rate $r$, the typical burst fluence $\mathcal{F}$, and the characteristic autocorrelation timescale $\tau$. These quantities can be interpreted in terms of the physical processes producing the gamma-ray emission.

One theoretical scenario that may give rise to a burst process of gamma-ray emission is magnetic reconnection, in which compact magnetized structures called plasmoids are produced that give rise to fast (minutes - days) gamma-ray flares \citep{Giannios2013}. Magnetic reconnection models based on particle-in-cell simulations predict that large, slow-moving plasmoids and much smaller but relativistic ones both produce flares with comparable fluence \citep{Petropoulou2016}. This finding suggests a motivation for the assumption applied in this work that all bursts have approximately the same fluence.

Simulations of plasmoid growth and mergers have found that in a single reconnection event, the bolometric light curve is dominated by flares from a few large plasmoids, along with a superposition of very fast variability contributed by smaller plasmoids \citep{Petropoulou2018, Christie2019}. Fast variability is especially evident in the high-energy gamma-ray light curve \citep{Acciari2020} in comparison to other wave bands, such as optical \citep{Zhang2022}. Simulations of magnetic reconnection have been shown to yield flux levels and variability patterns compatible with gamma-ray observations of flaring blazars, particularly FSRQs \citep{Meyer2021}. 

The autocorrelation timescale is most likely associated with longer-timescale variability. It may be directly connected to the same underlying physics as the burst process, such as very long duration flares produced by large plasmoids in a magnetic reconnection scenario \citep{Giannios2013}, or it may be indirectly related and associated with a different physical process. In particular, the variability of the jet emission may originate in processes in the accretion disk. In this case, $\tau$ may reflect one of several timescales associated with dynamical, thermal, or viscous processes in the disk \citep{Czerny2006}. Alternatively, quasi-periodic variability with a characteristic timescale of approximately 1 to 100 days could result from a ``striped jet'' characterized by magnetic fields of alternating polarity \citep{Giannios2019}. The variability of the light curve may be driven by different physical processes at different time scales. If so, $\tau$ may change when the light curve is examined using different choices of $\Delta T$. 

\subsection{Comparison to Previous Work}

The model presented in this paper shares several important similarities with the SDE proposed by \citet{Tavecchio2020}, along with a number of key differences. The model of \citet{Tavecchio2020} is motivated by the consideration of processes in a magnetically arrested accretion disk. In the model proposed in this work, the form of the flux distribution instead derives from the statistics of a burst process giving rise to the gamma-ray emission, although the long-term variations captured by the autoregressive structure may be considered to be related to processes in the accretion disk.

Both models predict AR(1) time structure and flux distributions with the form of an inverse gamma distribution. However, the models exhibit different behavior when the shape parameter of the distribution becomes small. In the autoregressive inverse gamma model, as $\alpha \to 0$, $\alpha_\mathrm{obs} \to 1$ and $\beta_\mathrm{obs} \to (1 - \phi)\beta_\mathrm{burst}$. Only a few time bins have nonzero flux, resulting in a light curve characteristic of a loner flare blazar. In the model of \citet{Tavecchio2020}, in the limiting case of $\lambda \to 0$ indicating the predominance of the stochastic term, the resulting flux distribution approaches a pure power-law distribution, $f(X) \propto X^{-2}$. 

In addition, the model proposed by \citet{Tavecchio2020} is represented as an SDE. The process is fundamentally continuous, although discrete approximations are numerically useful for generating simulations and fitting the model to data. The quality of the discrete approximation would be expected to increase for data that are increasingly finely binned in time (up to the effective time resolution of the instrument). On the other hand, the model proposed in this work is inherently discrete in character, as it is derived from a stochastic process of individual bursts. At short timescales, it would be expected to break down as the time structure of the bursts themselves begins to dominate the light curve. Mathematically, this discreteness comes about because the innovation is not Gaussian, which would be required to obtain a continuous Markov process \citep{Gillespie1996}.

In the model proposed by \citet{Tavecchio2020}, which is only supposed to be a simplified representation of the real dynamics, the stochastic term consists of Gaussian fluctuations multiplied by the normalized flux. As such, stochastic fluctuations can drive the flux below zero, especially when the flux is large. The requirement that the flux be greater than zero is required as an implicit additional constraint for deriving an inverse gamma flux distribution. In the model proposed in this work, the flux is inherently nonnegative, since the probability distributions have positive support.

The autoregressive inverse gamma model predicts a heavy-tailed flux distribution and approximately linear RMS-flux relation, similar to lognormal variability models based on multiplicative processes \citep{Uttley2005}. However, it explains these phenomena with an additive burst process modulated by long-term stochastic variability, without invoking multiplicative processes. In this way, the proposed model is more similar in its physical interpretation to shot-noise models that explain the light curve as resulting from the additive superposition of many overlapping bursts \citep[e.g.][]{Lehto1989, Tanihata2001}. However, there is a significant difference: because the bursts are unresolved, the flux distribution and power spectrum are determined by the statistics of the burst arrival process and the long-term stochastic variability, rather than by the shapes and amplitudes of the bursts themselves.

\citet{Duda2021} studied gamma-ray blazar variability using the log-stable family of probability distributions, which generalizes the lognormal distribution. Most of the blazars in their sample displayed variability consistent with lognormality; intriguingly, however, a subset of the sources had extremely heavy-tailed flux distributions leading to infinite variance. This finding of infinite variance is characteristic of the inverse gamma distribution with $\alpha_\mathrm{obs} < 2$. Interestingly, the sample of sources with infinite variance examined by \citet{Duda2021} contained both BL Lac objects and FSRQs.

Another model class that produces a heavy-tailed marginal distribution is that of generalized autoregressive conditional heteroskedasticity (GARCH) models, commonly used in finance for modeling time series of price fluctuations \citep{Engle1982, Bollerslev1986, Francq2019}. GARCH models describe white-noise processes in which the variance of the innovation changes over time, following an ARMA process. The time-varying variance is called the volatility.

The autoregressive inverse gamma model proposed in this work differs significantly from GARCH models, as it has a fixed innovation distribution. The model of \citet{Tavecchio2020}, which incorporates volatility in its stochastic term, is related to GARCH models. However, it differs from a standard first-order GARCH model by also having a mean-reverting autoregressive term.

The GARCH(1, 1) process has an inverse gamma marginal distribution of volatility, giving the resulting time series a Student’s \textit{t} marginal distribution \citep{Nelson1990}. \citet{Golan2013} proposed an alternative model for autocorrelated inverse gamma volatility in price fluctuations, in which exogenous events influencing trading activity play a similar role to the burst process considered in this work. The model proposed in this work goes beyond that of \citet{Golan2013} by allowing the number of bursts (or events) to fluctuate from time bin to time bin, including the possibility of no bursts (Sec.~\ref{sec:discretization}); enforcing an exact AR(1) autocorrelation structure; and having separate parameters modeling the underlying burst rate and autocorrelation timescale.

\subsection{Blazar Classification}

In the autoregressive inverse gamma model, the extreme variability typical of bright flaring FSRQs, such as that studied by \citet{Tavecchio2020} and \citet{Adams2022}, is associated with a small burst rate $r$. Relating high-energy gamma-ray variability to physically interpretable quantities may help characterize the differences observed between FSRQs and BL Lac objects, increasing our understanding of the relationship between these blazar classes. For example, \citet{Uemura2020} identified differences between the optical light curves of FSRQs and BL Lac objects by modeling them using Ornstein-Uhlenbeck processes. If FSRQs and BL Lac objects make up distinct classes rather than a continuous sequence, we might expect the burst parameters of these objects to belong to distinct clusters, with the opposite being true otherwise.

In some cases, blazar classifications are uncertain or disputed. For example, \citet{Padovani2019} have argued that the blazar TXS~0506+056, which has been associated with the detection of high-energy neutrinos, is an object intrinsically of the FSRQ type masquerading as a BL Lac object. Gamma-ray variability modeling may provide evidence to help us better understand these sources.

\section{Summary and Conclusions}\label{sec:conclusion}

Models of gamma-ray variability offer a critical tool for understanding the physical processes producing high-energy emission in blazars, potentially helping us characterize the relationship between FSRQs and BL Lac objects. In this work, we have proposed an autoregressive inverse gamma model of blazar variability that provides a simple framework for interpreting gamma-ray blazar variability on multiple timescales. In the proposed model, an inverse gamma flux distribution derives from an emission process in which gamma rays are produced in discrete bursts arriving as a Poisson process. Importantly, the bursts are taken to be individually unresolved within the time bins used in the analysis; only the average flux of each time bin is observed. Long-timescale stochastic variations are modeled by incorporating first-order autoregressive, or AR(1), structure.

The autoregressive inverse gamma model has three free parameters, representing the average burst rate, the burst fluence, and the autocorrelation timescale. These parameters can be interpreted in terms of physical quantities, such as by associating the bursts with plasmoid-powered flares in a magnetic reconnection scenario, for example. Furthermore, in the proposed model, flares and quiescent periods lasting from days to months are caused by random fluctuations in the arrival rate of the bursts. This intermediate-timescale activity naturally emerges from the interaction between two underlying physical processes, the short-timescale burst process and the long-timescale stochastic variations. Flaring and quiescent emission therefore have the same origin.

The autoregressive inverse gamma model yields simulated gamma-ray light curves consistent with multiple blazar source classes. The variability characteristics are controlled primarily by the burst rate. In particular, the use of parameters estimated from flare observations of the FSRQ 3C~279 yields a simulated light curve that demonstrates the variability behavior characteristic of that source, with bright flares separated by longer relatively quiescent periods, along with a similar overall flux normalization. Decreasing the burst rate generates a light curve consisting of isolated ``loner flares'', while increasing it produces light curves more closely resembling those of BL Lac objects. In this case, flaring activity is less prominent, and the flux distribution is approximately lognormal.

The proposed model is based on an emission process of discrete bursts, resulting in  novel predictions that differentiate it from previous work, including the model proposed by \citet{Tavecchio2020} that also features an inverse gamma flux distribution and autoregressive time structure. In particular, the fractional variability is predicted to decrease as a larger time binning is used, scaling as $F_\mathrm{var} \propto \Delta T^{-1/2}$.

Because the model has AR(1) structure, the PSD has the form of a power law with a low-frequency break. A high-frequency break may also occur, depending on the characteristics of the unresolved burst process.

\begin{acknowledgements}
    Thanks to Alberto Dom\'{i}nguez, Reshmi Mukherjee, Jeremy Perkins, Jeff Scargle, and Janeth Valverde for helpful discussions, comments, and suggestions on this work.
    A.B. is supported by the NASA Postdoctoral Program at Goddard Space Flight Center, administered by USRA and ORAU.
\end{acknowledgements}

\software{
NumPy \citep{Harris2020},
Matplotlib \citep{Hunter2007},
SciPy \citep{Virtanen2020}.
}

\appendix

\section{Derivation of an AR(1) Gamma Process with Poisson-distributed Burst Counts}\label{appendix:derivations}

In this section, we re-derive the gamma process proposed by \citet{Sim1990} to incorporate a more realistic model of burst counts. To do so, we replace the shape parameter $\alpha = r \Delta T$ that represents the average number of bursts in each bin with a Poisson distribution with parameter $\alpha$.

Since $\alpha$ is Poisson distributed, we have

\begin{equation}
    \alpha_\mathrm{bin} = 
    \begin{cases}
    0, & p = e^{-\alpha}\\
    l, & p = \frac{\alpha^l e^{-\alpha}}{l!},
    \end{cases}
\end{equation}

\noindent for natural numbers $l > 0$. A randomly chosen bin will with probability $e^{-\alpha}$ contain no bursts, and therefore generate no flux. It follows that time bins with observed flux will always have $\alpha_\mathrm{bin} > 0$, which is also a requirement for the process of \citet{Sim1990} to be well-defined. 

We therefore wish to construct an AR(1) process that switches between a zero and non-zero state (we will show later that in fact $F^{-1} = 0$ for bins with no bursts). To do so, we take inspiration from the Binary AR(1) process \citep{McKenzie1985}. Writing $F^{-1} \equiv X$ in the remainder of this section for brevity, we consider a Markov process with two states, $X = 0$ and $X \sim \Gamma(\alpha_\mathrm{obs}, \beta_\mathrm{obs})$, with transition probabilities for $X_i \to X_{i + 1}$ notated as

\begin{equation}
\begin{cases}
    p_{00}, & 0 \to 0\\
    p_{0\Gamma}, & 0 \to \Gamma(\alpha_\mathrm{obs}, \beta_\mathrm{obs})\\
    p_{\Gamma0}, & \Gamma(\alpha_\mathrm{obs}, \beta_\mathrm{obs}) \to 0\\
    p_{\Gamma\Gamma}, & \Gamma(\alpha_\mathrm{obs}, \beta_\mathrm{obs}) \to \Gamma(\alpha_\mathrm{obs}, \beta_\mathrm{obs}).
\end{cases}
\end{equation}

To ensure a self-consistent process, we have the relations

\begin{subequations}
\begin{align}
    p_{00} + p_{0\Gamma} &= 1 \label{eq:zero_consistency}\\
    p_{\Gamma0} + p_{\Gamma\Gamma} &= 1 \label{eq:gamma_consistency}\\
    p_0 + p_\Gamma &= 1,\label{eq:prob_consistency}
\end{align}
\end{subequations}

\noindent where $p_0 = e^{-\alpha}$ and $p_\Gamma = 1 - e^{-\alpha}$, as well as

\begin{subequations}
\begin{align}
    p_0 p_{00} + p_\Gamma p_{\Gamma0}  &= p_0 \label{eq:p0_consistency}\\
    p_0 p_{0\Gamma}+ p_\Gamma p_{\Gamma\Gamma} &= p_\Gamma \label{eq:pgamma_consistency}.
\end{align}
\end{subequations}

\begin{remark}
Let an \textit{isolated flare} be defined as a contiguous sequence of bins with $X \sim \Gamma$ surrounded by bins with $X = 0$ on both ends. The expected length of such a sequence is given by the expected number of time bins needed for one $\Gamma \to 0$ transition to occur, starting from an initial time bin with $X \sim \Gamma$. The average isolated flare length in units of number of time bins is therefore

\begin{equation}
    L = \frac{1}{p_{\Gamma 0}}.
\end{equation}
\end{remark}

\begin{remark}
Let the \textit{effective duration} of a time bin with $X \sim \Gamma$ be defined as the total duration encompassing that time bin and any subsequent time bins with $X = 0$, until the next bin with $X \sim \Gamma$ is reached. The expected length of a sequence of bins with $X = 0$ is $1/p_{0\Gamma}$. Making use of this fact and Eqs.~\ref{eq:pgamma_consistency} and \ref{eq:prob_consistency}, the expected effective duration is

\begin{equation}
    N_\mathrm{eff} = \Delta T \left(1 + \frac{p_{\Gamma 0}}{p_{0 \Gamma}} \right) = \Delta T \left( 1 + \frac{p_0}{p_\Gamma} \right) = \frac{\Delta T}{p_\Gamma} = \frac{\Delta T}{1 - e^{-\alpha}}\label{eq:nominal_duration}.
\end{equation}
\end{remark}

We now focus on the case $\alpha_\mathrm{bin} > 0$, that is, the process that takes place when the transition $\Gamma \to \Gamma$ occurs, and will return later to finish considering the remaining cases. Renormalizing the probability distribution of $\alpha_\mathrm{bin}$ for $\alpha_\mathrm{bin} > 0$ yields

\begin{equation}
    \alpha_\mathrm{bin,obs} = 
    \begin{cases}
    l, & p = \frac{1}{1 - e^{-\alpha}} \frac{\alpha^l e^{-\alpha}}{l!}.
    \end{cases}
\end{equation}

For a time bin with $\alpha_\mathrm{bin,obs} = l$, the corresponding observed rate is

\begin{equation}
    \beta_\mathrm{bin,obs} = \frac{\Delta T}{N_\mathrm{eff}} \beta_\mathrm{burst} l = \frac{1 - e^{-\alpha}}{\alpha} \beta l,
\end{equation}

\noindent where two modifications have been incorporated. First, the flux is averaged over $l$ bursts rather than $\alpha$ bursts. Second, the average flux is reduced by a ``leakage'' factor of $\Delta T/N_\mathrm{eff}$ because the fluence produced in a time bin is supposed to be averaged over the burst interarrival time, which is the entire effective duration $N_\mathrm{eff}$, but the flux is only collected within that bin itself, of duration $\Delta T$.

The process mean $\mathbb{E}(X)$ and variance $\mathrm{Var}(X)$ can be calculated as follows. From the law of total expectation, we have

\begin{equation}\label{eq:mu}
\begin{split}
    \mathbb{E}(X) &= \mathbb{E}(X_{i + 1})\\
    &= \mathbb{E}_{X_i}\left[ \sum_{k = 0}^{\infty} \sum_{l = 1}^{\infty} \frac{(\phi \frac{1 - e^{-\alpha}}{\alpha} \beta l X_i)^k e^{-\phi \frac{1 - e^{-\alpha}}{\alpha} \beta l X_i}}{k!} \frac{1}{1 - e^{-\alpha}} \frac{\alpha^l e^{-\alpha}}{l!} \mathbb{E}_{X_{i + 1}}[\Gamma(X_{i + 1}; k + l, \frac{1 - e^{-\alpha}}{\alpha} \beta l)] \right]\\
    &= \frac{1}{(1 - e^{-\alpha})^2} \frac{\alpha}{\beta} \mathbb{E}_{X_i}\left[ \sum_{k = 0}^{\infty} \sum_{l = 1}^{\infty}  \frac{(\phi \frac{1 - e^{-\alpha}}{\alpha} \beta l X_i)^k e^{-\phi \frac{1 - e^{-\alpha}}{\alpha} \beta l X_i}}{k!}  \frac{\alpha^l e^{-\alpha}}{l!} \left( \frac{k}{l} + 1 \right) \right]\\
    &= \frac{1}{(1 - e^{-\alpha})^2} \frac{\alpha}{\beta} \mathbb{E}_{X_i}\left[ \sum_{l = 1}^{\infty} \frac{\alpha^l e^{-\alpha}}{l!} \left( \phi \frac{1 - e^{-\alpha}}{\alpha} \beta X_i + 1 \right) \right]\\
    &= \frac{1}{1 - e^{-\alpha}} \frac{\alpha}{\beta} \mathbb{E}_{X_i}\left[ \phi \frac{1 - e^{-\alpha}}{\alpha} \beta X_i + 1 \right]\\
    &= \phi \mathbb{E}(X) + \frac{1}{1 - e^{-\alpha}}\frac{\alpha}{\beta}\\
    &= \frac{1}{1 - \phi} \frac{1}{1 - e^{-\alpha}} \frac{\alpha}{\beta},
\end{split}
\end{equation}

\noindent where we have made use of the fact that for a stationary process, $\mathbb{E}(X_{i + 1}) = \mathbb{E}(X_i) = \mathbb{E}(X)$.

From the law of total variance, we have

\begin{equation}\label{eq:sigma2_1}
\begin{split}
    \mathrm{Var}(X) &= \mathbb{E}(\mathrm{Var}(X_{i + 1} | X_i)) + \mathrm{Var}(\mathbb{E}(X_{i + 1} | X_i))\\
    &= \mathbb{E}(\mathrm{Var}(X_{i + 1} | X_i)) + \mathrm{Var}\left(\phi X_i + \frac{1}{1 - e^{-\alpha}}\frac{\alpha}{\beta}\right)\\
    &= \mathbb{E}(\mathrm{Var}(X_{i + 1} | X_i)) + \phi^2 \mathrm{Var}(X)\\
    &= \frac{1}{1 - \phi^2}\mathbb{E}(\mathrm{Var}(X_{i + 1} | X_i)).
\end{split}
\end{equation}

For a mixture of distributions $f_i$ with means $\mu_i$, variances $\sigma^2_i$, and weights $p_i$, the variance of the mixture $f = \sum p_i f_i$ is given by

\begin{equation}\label{eq:variance_mixture}
    \mathrm{Var}(f) = \sum_i p_i \left( \sigma_i^2 + \mu_i^2 \right) - \left(\sum_i p_i \mu_i \right)^2.
\end{equation}

Using Eq.~\ref{eq:variance_mixture}, we can therefore write

\begin{equation}\label{eq:sigma2_2}
\begin{split}
    \begin{aligned}\mathrm{Var}(X_{i + 1} | X_i) = \: \\ \end{aligned} &\overbrace{\begin{aligned}\sum_{k = 0}^{\infty} \sum_{l = 1}^{\infty} \frac{(\phi \frac{1 - e^{-\alpha}}{\alpha} \beta l X_i)^k e^{-\phi \frac{1 - e^{-\alpha}}{\alpha} \beta l X_i}}{k!} \frac{1}{1 - e^{-\alpha}} \frac{\alpha^l e^{-\alpha}}{l!} \biggl[ &\mathrm{Var}(\Gamma(X_{i + 1}; k + l, \frac{1 - e^{-\alpha}}{\alpha} \beta l)) \\ &+ \mathbb{E}(\Gamma(X_{i + 1}; k + l, \frac{1 - e^{-\alpha}}{\alpha} \beta l))^2 \biggr] \end{aligned}}^{(A)}\\
    &- \underbrace{\left(\sum_{k = 0}^{\infty} \sum_{l = 1}^{\infty} \frac{(\phi \frac{1 - e^{-\alpha}}{\alpha} \beta l X_i)^k e^{-\phi \frac{1 - e^{-\alpha}}{\alpha} \beta l X_i}}{k!} \frac{1}{1 - e^{-\alpha}} \frac{\alpha^l e^{-\alpha}}{l!} \mathbb{E}(\Gamma(X_{i + 1}; k + l, \frac{1 - e^{-\alpha}}{\alpha} \beta l)) \right)^2}_{(B)},
\end{split}
\end{equation}

\noindent where

\begin{equation}\label{eq:a}
\begin{split}
    (A) &= \frac{1}{1 - e^{-\alpha}} \sum_{k = 0}^{\infty} \sum_{l = 1}^{\infty} \frac{(\phi \frac{1 - e^{-\alpha}}{\alpha} \beta l X_i)^k e^{-\phi \frac{1 - e^{-\alpha}}{\alpha} \beta l X_i}}{k!} \frac{\alpha^l e^{-\alpha}}{l!} \frac{(k + l) + (k + l)^2}{\left(\frac{1 - e^{-\alpha}}{\alpha} \beta l \right)^2 }\\
    &= \frac{1}{1 - e^{-\alpha}} \frac{\alpha}{\beta^2} \frac{\alpha}{(1 - e^{-\alpha})^2} \sum_{k = 0}^{\infty} \sum_{l = 1}^{\infty} \frac{(\phi \frac{1 - e^{-\alpha}}{\alpha} \beta l X_i)^k e^{-\phi \frac{1 - e^{-\alpha}}{\alpha} \beta l X_i}}{k!} \frac{\alpha^l e^{-\alpha}}{l!} \frac{k^2 + 2kl + l^2 + k + l}{l^2}\\
    &= \frac{1}{1 - e^{-\alpha}} \frac{\alpha}{\beta^2} \frac{\alpha}{(1 - e^{-\alpha})^2} \sum_{l = 1}^{\infty} \frac{\alpha^l e^{-\alpha}}{l!} \left[ \left(\phi \frac{1 - e^{-\alpha}}{\alpha} \beta X_i\right)^2 + 2\phi \frac{1 - e^{-\alpha}}{\alpha} \beta X_i + 1 + \frac{2\phi \frac{1 - e^{-\alpha}}{\alpha} \beta X_i + 1}{l} \right]\\
    &= (\phi X_i)^2 + \frac{2\phi\alpha\beta X_i}{1 - e^{-\alpha}} + \left(\frac{\alpha}{1 - e^{-\alpha}}\right)^2 + \frac{1}{1 - e^{-\alpha}} \frac{\alpha}{\beta^2} \left( 2\phi \frac{1 - e^{-\alpha}}{\alpha} \beta X_i + 1 \right) A^{-1}(\alpha),
\end{split}
\end{equation}

\noindent with

\begin{equation}
    A(\alpha) = \left( \frac{\alpha}{(1 - e^{-\alpha})^2} \sum_{l = 1}^{\infty} \frac{\alpha^l e^{-\alpha}}{l!} \frac{1}{l} \right)^{-1},
\end{equation}

\noindent and

\begin{equation}\label{eq:b}
\begin{split}
    (B) &= \frac{1}{(1 - e^{-\alpha})^4} \frac{\alpha^2}{\beta^2}\left[\sum_{k = 0}^{\infty} \sum_{l = 1}^{\infty} \frac{(\phi \frac{1 - e^{-\alpha}}{\alpha} \beta l X_i)^k e^{-\phi \frac{1 - e^{-\alpha}}{\alpha} \beta l X_i}}{k!} \frac{1}{1 - e^{-\alpha}} \frac{\alpha^l e^{-\alpha}}{l!} \left(\frac{k}{l} + 1 \right) \right]^2\\
    &= \left(\phi X_i + \frac{1}{1 - e^{-\alpha}}\frac{\alpha}{\beta} \right)^2 \mathrm{(From~Eq.~\ref{eq:mu})}\\
    &= (\phi X_i)^2 + \frac{2\phi\alpha\beta X_i}{1 - e^{-\alpha}} + \left(\frac{\alpha}{1 - e^{-\alpha}}\right)^2.
\end{split}
\end{equation}

Combining Eqs.~\ref{eq:sigma2_1}, \ref{eq:sigma2_2}, \ref{eq:a}, and \ref{eq:b} and making use of Eq.~\ref{eq:mu} in the second step, 

\begin{equation}
\begin{split}
    \mathrm{Var}(X) &= \frac{1}{1 - \phi^2}\frac{1}{1 - e^{-\alpha}} \frac{\alpha}{\beta^2}A^{-1}(\alpha)\mathbb{E}\left[2\phi\frac{1 - e^{-\alpha}}{\alpha} \beta X_i + 1 \right]\\
    &= \frac{1}{1 - \phi^2}\frac{1}{1 - e^{-\alpha}} \frac{\alpha}{\beta^2}A^{-1}(\alpha)\left(2\phi\frac{1 - e^{-\alpha}}{\alpha} \beta \frac{1}{1 - \phi} \frac{1}{1 - e^{-\alpha}} \frac{\alpha}{\beta} + 1 \right)\\
    &= \frac{1}{1 - \phi^2}\frac{1}{1 - e^{-\alpha}} \frac{\alpha}{\beta^2}A^{-1}(\alpha)\left(\frac{2\phi}{1 - \phi} + 1 \right)\\
    &= \frac{1}{(1 - \phi)^2}\frac{1}{1 - e^{-\alpha}} \frac{\alpha}{\beta^2}A^{-1}(\alpha).
\end{split}
\end{equation}

The resulting mixture of gamma distributions does not have the exact form of a gamma distribution, but it can be reasonably well approximated as one and approaches an exact gamma distribution in the limits of $\alpha \to 0$ and $\alpha \to \infty$. The best-fit parameters of the mixture to a gamma distribution can be estimated using the method of moments,

\begin{equation}
    \alpha_\mathrm{obs} = \frac{\mathbb{E}(X)^2}{\mathrm{Var}(X)} = \frac{A(\alpha)}{1 - e^{-\alpha}}\alpha,
\end{equation}

\noindent and

\begin{equation}
    \beta_\mathrm{obs} = \frac{\mathbb{E}(X)}{\mathrm{Var}(X)} = A(\alpha)(1 - \phi)\beta.
\end{equation}

We can now validate our assumption that $F^{-1} = 0$ for bins with no bursts. The mean of the average process $X_\mathrm{avg} \sim \Gamma(\alpha, \beta)$ is

\begin{equation}\label{eq:mu_total}
    \mathbb{E}(X_\mathrm{avg}) = \frac{1}{1 - \phi}\frac{\alpha}{\beta},
\end{equation}

\noindent which should be equal to the mean of the combined process,

\begin{equation}
\begin{split}
    \mathbb{E}(X_\mathrm{avg}) &= e^{-\alpha}\mathbb{E}(X_0) + (1 - e^{-\alpha})\mathbb{E}(X_\mathrm{obs})\\
    \frac{1}{1 - \phi}\frac{\alpha}{\beta} &= e^{-\alpha}\mathbb{E}(X_0) + (1 - e^{-\alpha})\frac{1}{1 - \phi} \frac{1}{1 - e^{-\alpha}} \frac{\alpha}{\beta},
\end{split}
\end{equation}

\noindent from which we obtain $\mathbb{E}(X_0) = 0$. Since $F^{-1} \geq 0$, it follows that $F^{-1} = 0$ for all bins with no bursts. The flux for bins with no bursts is therefore undefined.

We now return to the full process, considering both bins with and without bursts. To impose on the process AR(1) correlation structure with autocorrelation constant $\phi$, we use the condition 

\begin{equation}\label{eq:phi_total}
    \phi = \frac{\mathrm{Cov}(X_{i + 1}, X_i)}{\mathrm{Var}(X)} = \frac{\mathbb{E}(X_{i + 1} X_i) - \mathbb{E}(X)^2}{\mathrm{Var}(X)}.
\end{equation}

From the law of total expectation,

\begin{equation}\label{eq:cond_expectation_total}
\begin{split}
    \mathbb{E}(X_{i + 1} X_i) &= p_0 p_{00} \cdot 0^2 + (p_0 p_{0\Gamma} + p_\Gamma p_{\Gamma 0}) \cdot 0 \cdot \mathbb{E}(X_\Gamma) + p_\Gamma p_{\Gamma \Gamma} \mathbb{E}(X_{\Gamma, i}, X_{\Gamma, i + 1})\\
    &= p_\Gamma p_{\Gamma \Gamma} \mathbb{E}(X_{\Gamma, i}, X_{\Gamma, i + 1})\\
    &= p_\Gamma p_{\Gamma \Gamma} \left[\phi \mathrm{Var}(X_\Gamma) + \mathbb{E}(X_\Gamma)^2 \right]\\
    &= p_{\Gamma \Gamma} \frac{1}{(1 - \phi)^2} \frac{\alpha^2}{\beta^2} \left( \frac{\phi}{\alpha A(\alpha)} + \frac{1}{1 - e^{-\alpha}} \right)\\
\end{split}
\end{equation}

Using Eq.~\ref{eq:variance_mixture},

\begin{equation}\label{eq:variance_total}
\begin{split}
    \mathrm{Var}(X) &= p_0 \left( \mathrm{Var}(X_0) + \mathbb{E}(X_0)^2 \right) + p_\Gamma \left( \mathrm{Var}(X_\Gamma) + \mathbb{E}(X_\Gamma)^2 \right) - \left( p_0 \mathbb{E}(X_0) + p_\Gamma \mathbb{E}(X_\Gamma) \right)^2\\
    &= p_\Gamma \left[ \mathrm{Var}(X_\Gamma) + \mathbb{E}(X_\Gamma)^2 - p_\Gamma \mathbb{E}(X_\Gamma)^2 \right]\\
    &= \frac{1}{(1 - \phi)^2} \frac{\alpha^2}{\beta^2} \left( \frac{1}{\alpha A(\alpha)}  + \frac{e^{-\alpha}}{1 - e^{-\alpha}} \right).
\end{split}
\end{equation}

Combining Eqs.~\ref{eq:phi_total}, \ref{eq:cond_expectation_total}, \ref{eq:mu_total}, and \ref{eq:variance_total} and simplifying yields

\begin{equation}
    p_{\Gamma\Gamma} = 1 - \frac{(1 - \phi)e^{-\alpha}}{1 + \frac{\phi}{\alpha}A^{-1}(\alpha)(1 - e^{-\alpha})} = 1 - p(\phi, \alpha)e^{-\alpha},
\end{equation}

\noindent where

\begin{equation}
    p(\phi, \alpha) =  \frac{1 - \phi}{1 + \frac{\phi}{\alpha}A^{-1}(\alpha)(1 - e^{-\alpha})}.
\end{equation}

From Eqs.~\ref{eq:zero_consistency} and \ref{eq:gamma_consistency}, and Eq.~\ref{eq:p0_consistency} or \ref{eq:pgamma_consistency}, we obtain the full set of transition probabilities,

\begin{equation}
\begin{split}
    p_{00} &= 1 - p(\phi, \alpha)(1 - e^{-\alpha})\\
    p_{0\Gamma} &= p(\phi, \alpha)(1 - e^{-\alpha})\\
    p_{\Gamma0} &= p(\phi, \alpha)e^{-\alpha}\\
    p_{\Gamma\Gamma} &= 1 - p(\phi, \alpha)e^{-\alpha}.
\end{split}
\end{equation}

\bibliography{bibliography}

\end{document}